\pdfoutput=1
\documentclass[prb,twocolumn,showpacs,showkeys,preprintnumbers,amsmath,amssymb]{revtex4}
\usepackage{graphicx}
\usepackage[T1]{fontenc}
\usepackage{amssymb,amsmath}
\usepackage{dcolumn}
\usepackage{bm}

\begin{document}

\title{Calculation of the Capacitances of Conductors ---
Perspectives for the Optimization of Electronic Devices}
\author{Thilo Kopp and Jochen Mannhart}
\affiliation{Center for Electronic Correlations and Magnetism, Institute of Physics,
                      Universit\"at Augsburg, D-86135 Augsburg, Germany}
\date{\today}

\begin{abstract}
The equation describing the capacitance of capacitors is determined. It is shown that by optimizing 
the material of the conducting electrodes, the capacitance of capacitors reaching the quantum regime
can be substantially enhanced or reduced. Dielectric capacitors with negative total capacitances are suggested and their properties analyzed. Resulting perspectives to enhance the performance of electronic devices are discussed. 

\end{abstract}
\pacs{71.27.+a,73.20.-r,73.21.-b,73.40.-c}
\keywords{negative capacitance, strongly correlated electrons, multiferroics}

\maketitle

\section{Introduction}
As the characteristic properties of electronic circuits reaches quantum scales, new possibilities emerge for the realization of novel, quantum electronic devices which exploit the quantum nature of solids.
Here we show that by using quantum effects, the capacitance of electronic devices can be optimized to a great extent, which is a key to the further miniaturization of electronic components.~\cite{Meindl01,Wilk01,Taur02,Robertson06}
While in MOSFETs, for example, large gate capacitances are needed for operation at small gate voltages,
their gate insulators are required to have an equivalent thickness of at least 1~nm to keep the gate currents at acceptably small values. To meet both requirements, high-$\kappa$ gate insulators,
consisting for example of HfO$_2$ based compounds, have been introduced into the gate stack of MOSFETs used in microprocessors.

As the sizes of the devices approach the nanometer length scale, quantum mechanical phemonena 
become increasingly relevant. The well-known classical models of semiconducting devices and the corresponding scaling laws break down. Ballistic transport processes, for example, gain importance, as does the statistical distribution of the dopant atoms in the small drain-source channels.

Quantum effects also alter the capacitance of large, macroscopic capacitors that reach the quantum regime because their dielectrics become thin. As we show, for such devices the capacitance comprises not only the well-known geometric capacitance, but four more capacitance
terms. These originate from the kinetic energy of the electrons, from their exchange energy, their correlation energy, and the electron-phonon coupling energy. 

In this paper, we
present the general equation for the capacitance of capacitors that considers the quantum mechanical energies of the electronic systems involved. This equation applies to two-plate capacitors, also in case their dielectric is ultrathin.
It is shown that for given $\epsilon_r$ and $A$ the capacitance of capacitors reaching the quantum regime may differ considerably from the well-known value 
\begin{equation}
\label{capacity}
C=\frac{\epsilon_0 \epsilon_r A}{d}
\end{equation} 
In fact, the capacitance may be even negative. Here $\epsilon_0$ is the dielectric constant of the vacuum, $\epsilon_r$ is the dielectric constant of the dielectric material between the two electrodes of area $A$, and $d$ is the thickness of the dielectric.

It is specifically shown that capacitors can be fabricated that do not follow Eq.~(\ref{capacity}). By altering the conducting materials of the electrodes and by varying the geometry of the capacitor in novel ways the capacitance of such capacitors can be greatly enhanced or lowered as compared to Eq.~(\ref{capacity}). Electrode materials of interest include materials with strong electronic correlations such as transition metal oxides, two-dimensional electron gases as generated at interfaces in oxides or semiconductors, metals with small carrier densities, organic conductors, and materials such as graphene. We point out that these quantum phenomena and their technical implementation provide a possible strategy to control large or small capacitances by the electronic properties of the conducting electrodes --- as opposed to the standard search for low or high-$\kappa$ dielectrics~\cite{Schlom08}.

It is common knowledge that classical electrostatics~\cite{Jackson} determines the capacitances. 
In the textbook case of a capacitor with two electrodes, the capacitance is characterized by 
\begin{equation}
\label{charge}
Q = CV 
\end{equation}
where $Q$ is the charge on a capacitor electrode, $C$ is the capacitance, and $V$ is the voltage between the electrodes.
In more general terms, for multiple conductors $(i=1,2,\ldots,n)$, the capacitance is characterized by the 
corresponding static response relations $ Q_i=\sum_j C_{ij} V_j $,
where $V_i$ and $Q_i$ are the potential and the total charge of conductor $i$, respectively. The coefficients $C_{ij}$
define the capacitance matrix. 
According to the classical approach, the coefficients of a capacitor consisting of 
two infinitely extended parallel metallic plates
are $C_{11}=C_{22}=-C_{12}=-C_{21}\equiv C$  (Eq.~(\ref{capacity})). 
Here, fringe field effects have been neglected. In this classical description, the capacitance is not influenced by the electronic properties of the conducting plates. For a two-plate capacitor, the geometric capacitance is entirely determined by the effective distance $d^\star=d/\epsilon_r$.

It is apparent that Eq.~(\ref{capacity}) is not very useful for the design of ultra-small devices, where, for example, single electron effects play an essential role (see, e.g., Ref.~\onlinecite{Datta}). The calculation of the electronic properties of such devices is typically based on the direct calculation of quantum mechanical energies.

Quantum mechanical effects, however, are also of importance for large devices with macroscopic lateral dimensions. The conductors of electrodes can screen, for example, the electric field only over non-vanishing lengths such as the Thomas Fermi screening length. This effect obviously enhances the effective dielectric thickness $d^\star$. 
The enhancement of $d^\star$ has first been observed by Mead~\cite{Mead61}, confirmed by Hebard {\it et al.}~\cite{Hebard87} for a known thickness of the dielectric,  and deduced\cite{Ku64,Simmons65}
from the charge distribution profile in the electrodes which is controlled by the kinetic energy of the charge carriers~ (see the detailed discussion and evaluations in Refs.~\onlinecite{Stern72}).
A framework to calculate the capacitance coefficients $C_{ij}$ that considers the field penetration into the conductor
has been formulated by  B{\"u}ttiker.~\cite{Buettiker93} He evaluated the space-dependent Lindhard function. The field penetration into the capacitor electrodes introduces corrections to the geometrical capacitance:~\cite{Buettiker93} 
in the case of space independent density of states (DOS) right up to the surface 
$ A/C= (d/\epsilon_r + \lambda_1^{\rm (TF)} +\lambda_2^{\rm (TF)})/\epsilon_0$ where $\lambda_{1,2}^{\rm (TF)}$ are the Thomas-Fermi screening lengths of the two metal plates. For a capacitor with two-dimensional metallic plates, $\lambda_1^{\rm (TF)} +\lambda_2^{\rm (TF)}$ is replaced~\cite{Luryi88,Buettiker93} by $a_B/2$ where $a_B=4\pi\epsilon_0\hbar^2/(me^2)$ is the Bohr radius. 

Lang and Kohn\cite{Lang73} introduced the center of mass  positions $x_0^{(1,2)}$ of  the induced surface charge density at the electrodes and related the distance between the center of mass positions 
to the  capacitance of a parallel plate capacitor in the vacuum ($\epsilon_r=1$): $A/C=(x_0^{(1)}-x_0^{(2)})/\epsilon_0$. They calculated $x_0^{(i)}$  within density-functional theory (DFT)\cite{Lang70} for a metal with a uniform distribution of positive background charge (jellium model). They determined the center of mass positions to be outside of the surface of the electrode (given by the edge of the positive background charge at $x_b^{(1,2)}$) so that the capacitance is increased with respect to its geometric value $\epsilon_0 A/ (x_b^{(1)}-x_b^{(2)})$. The information on the electrodes is encoded in the distances $\delta x^{(1,2)} = x_0^{(1,2)}- x_b^{(1,2)}$ which generate interface capacitances serially connected to the geometric capacitance.

For a macroscopic $d^\star$, the corrections arising from the Thomas-Fermi lengths or  $\delta x^{(1,2)}$ are tiny, because they represent  small, quantum mechanical length scales.
However, exceeding these effects described, quantum phenomena are of fundamental relevance in controlling the capacitances of electronic circuits as soon as the effective thickness of their dielectric layers $d^\star$ approaches quantum mechanical lengths such as $\epsilon_r a_B$. Current devices like modern MOSFETs are close to this transition, and quantum phenomena need to be considered in their design. As we show, this even holds for capacitors of macroscopic lateral dimensions and macroscopic numbers of charge carriers, if their dielectric layers are sufficiently thin. It is hereby particularly interesting that also correlation effects in the electron systems of the conducting electrodes may alter the capacitances in unique ways. We therefore propose to use these quantum phenomena as a possible tool to optimize devices such as MOSFETs. Desired large or small capacitances can potentially be attained by tuning the electronic properties of the conducting electrodes --- in addition to the standard approach of using low- or high-$\kappa$ dielectrics and small or large barrier thicknesses.

\section{The Composition of the Capacitance of Macroscopic Capacitors}

To determine the equation that describes the capacitance of capacitors, we start by considering the energy functional for the electronic ground state of a given capacitor. For clarity, we restrict our evaluation to capacitors with two parallel plates and neglect effects arising from finite temperatures and from fringe fields. The plates are considered to be so far apart that their electronic wave functions do not overlap. Tunneling and exchange contributions between the plates are therefore negligible.\cite{commentT} The ground state energy as a functional of the electronic densities may be decomposed as:~\cite{Mahan}

\begin{equation}
\label{energy}
E=E_{\rm H} + \sum_i E_{{\rm kin},i} +  \sum_i E_{{\rm x},i} +  \sum_i E_{{\rm c},i} +  \sum_i E_{{\rm ext},i}
\end{equation}
The sums are used to describe the energies of the two possibly different metallic plates. In capacitors for which surfaces are relevant, the electronic density $n_i({\bm r})$ is a function of the space coordinate ${\bm r}$.
The energy  $E[n_i({\bm r})]$  is then a functional of the densities on the two electrodes. 

The Hartree term $E_{\rm H}$ in Eq.~(\ref{energy}) includes all direct Coulomb interactions between the electrons of the conduction bands; these long-range Coulomb potentials also include the electrostatic interaction between the plates.  The other energies, which are of pure electronic origin, are not caused by inter-plate interactions, but rather are local attributes of each plate. These energies are the kinetic energy of the electrons $E_{{\rm kin},i}$, the exchange or Fock energy $E_{{\rm x},i}$, and the correlation energy $E_{{\rm c},i}$, which represents the electron-electron interaction energies beyond the Hartree-Fock description. 

The external energy $E_{{\rm ext},i}$ is the energy of the conduction electrons in an external potential; it most prominently includes the interaction energy between the electrons and the background charge that is caused by the nuclei. Because this Coulomb interaction between electrons and nuclei is long-ranged, it diverges with increasing area $A$; this divergence is canceled by a similar diverging term of opposite sign, which results from the electron-electron interaction in the Hartree contribution.~\cite{commentV} Inhomogeneities such as surface potentials or potential wells are also introduced through $E_{{\rm ext},i}$.

To treat the electronic exchange effects, all electron-electron interactions, including those with the electrons of the ionic core states, need to be contained in the Hartree-Fock and correlation terms. In effective models, however, the exchange contribution of ionic core states with conduction electrons is often neglected, an approximation that allows to include the  occupation of these electronic core states in a pseudopotential or in a suitably defined ionic core charge. In those cases, it is useful to introduce an effective dielectric constant $\epsilon_{{\rm eff},i}$ into the electronic interaction terms  which accounts for the polarizability of the ionic cores.

In case the electron systems on the two parallel plates are homogeneous,  i.e.\ if they don't reside on lattices, \cite{comment0} $E_{\rm H} +\sum_i E_{{\rm ext},i}=Q^2 d/(2 \epsilon_0 \epsilon_r A)$. 
In such a situation, 
$d$ is given by the distance between
the boundaries of the uniform positive background charge of the electrodes.
This energy term matches the standard equation for the energy of the capacitor, 

\begin{equation}
\label{geom-energy}
E_{\rm Coul}=Q^2/(2C_{\rm geom})
\end{equation}
with the geometric capacitance
$C_{\rm geom}={\epsilon_0 \epsilon_r A}/d$.
Yet, also the other three terms in the energy functional generate capacitances. These capacitances result from distinct quantum phenomena. 

To derive the total capacitance $C$ for $Q\rightarrow 0$, we expand the energy functional to second order in $Q$.  If the two plates differ in their physical properties, also a linear contribution in $Q$ arises, which originates from the difference in the work functions of the electrode materials. The work function $\Phi$, defined as the minimum energy necessary to extract an electron, is composed of the chemical potential of the infinite system (relative to the electrostatic potential in this system) and the electrostatic potential  energy difference across the respective surface of the electrode 
(for details and the evaluation see, e.g., Refs.~\onlinecite{Bardeen36} and \onlinecite{Lang70}). The contribution 
to the work function from the chemical potential is generated by all energy terms. The electrostatic potential difference across the surface, the so-called electrostatic dipole barrier, is caused by the external energy. A difference $\Delta\Phi$ of the work functions of the surfaces of the opposing electrodes generates a contact voltage, so that associated surface charges $\pm Q_{\rm s}$ are induced. This  electronic charge reconstruction compensates the work function mismatch. The differential capacitance can be evaluated by expanding the energy functional around the charge-reconstructed states of the electrodes.

\begin{figure}[t]
\centering
\includegraphics[width=0.99\columnwidth]{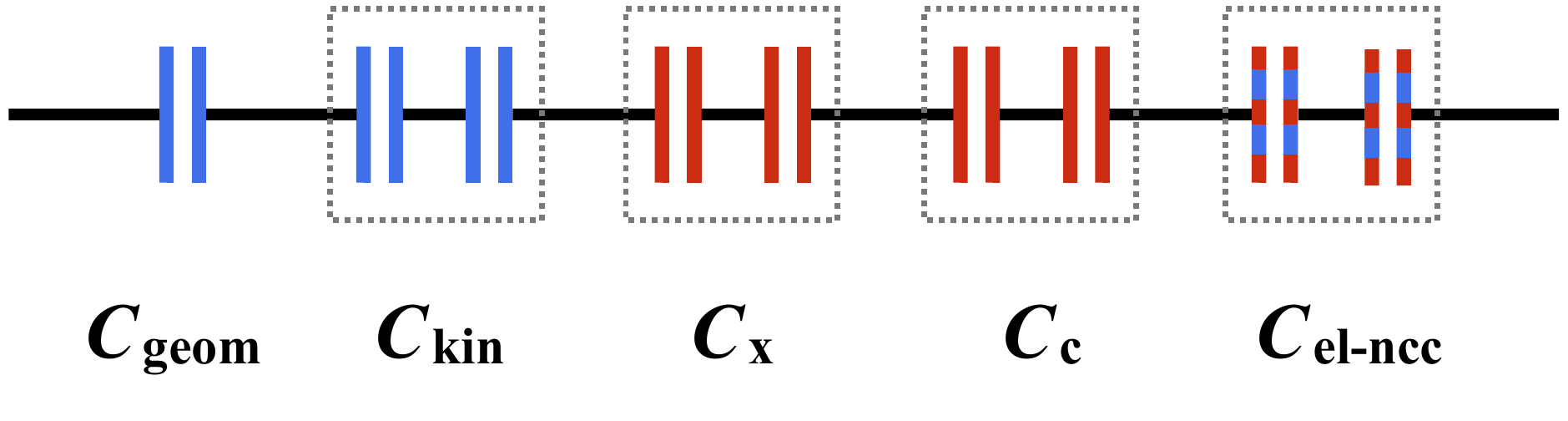}
\caption
{Equivalent circuit diagram of a two-plate capacitor according to Eq.~(\ref{capacitance}). The capacitor consists of a serial connection of the geometric capacitance and the quantum capacitances of each of the two plates. The capacitance results from the Hartree energy ($\propto 1/C_{\rm geom}$), 
the kinetic energy ($\propto 1/C_{\rm kin}$), the exchange energy ($\propto 1/C_{\rm x}$), 
the correlation energy ($\propto 1/C_{\rm c}$), and the short range interaction energy of charge carriers
with all ``non charge-carrier'' (ncc) degrees of freedom ($\propto 1/C_{\rm el-ncc}$). Positive and negative capacitances are drawn in both blue and red, respectively. The capacitance $C_{\rm el-ncc}$ is presented in blue and red, because the sign of the capacitance depends on the microscopic model for the coupling of the ncc degrees freedom to the charge carriers;
for phonons it is expected to be negative.
}
\label{SerialCap}
\end{figure}

Focussing first on the capacitance 
\begin{equation}
\label{diff-capacitance}
C=Q/V
\end{equation}
in linear response, i.e., for $V\rightarrow 0$ (or, for finite $Q_{\rm s}=-C\Delta\Phi/e$, 
on~\cite{commentcap} $C=(Q-Q_{\rm s})/V$),
we collect the quadratic terms and identify the inverse capacitances according to the coefficients of $Q^2/2$:
\begin{eqnarray}
\label{capacitance}
1/C =  1/C_{\rm geom} && +  \sum_i 1/C_{{\rm kin},i} +  \sum_i 1/C_{{\rm x},i}\nonumber \\
&& +  \sum_i 1/C_{{\rm c},i} +  \sum_i 1/C_{{\rm el-ncc},i}
\end{eqnarray}
Eq.~(\ref{capacity}) is replaced by this general equation, which reveals that the total capacitance is caused by a series connection of the capacitances that result from the additive terms in the energy functional (Fig.~\ref{SerialCap}). Although Eq.~(\ref{capacity}) seems to imply that the total capacitance is always smaller than the geometrical capacitance, the total capacitance can, in fact, well exceed the geometrical one, because $C_{\rm x}$ and $C_{\rm c}$ may have large negative values.

Note that the external energy $E_{{\rm ext},i}$ consists of
a long range Coulomb term, which compensates the diverging term of the purely electronic Hartree energy, and a remaining short range contribution. 
The inverse geometric capacitance $1/C_{\rm geom}$ is generically defined as the $Q^2$-coefficient from the Hartree energy and the static long-range part of the external energy.
The short range part of  $E_{{\rm ext},i}$ contribution includes the interaction between the charge carriers and all ``non-charge carrier degrees of freedom'' (ncc) such as phonons or local potentials. Examples for the latter are the static potentials of the ionic cores and the confining potentials in semiconductor heterojunctions. These  contributions generate the $1/C_{{\rm el-ncc}}$ term in Eq.~(\ref{capacitance}). In this paper we will not investigate $C_{{\rm el-ncc}}$  in more detail.  

Eventually, to obtain the voltage dependent capacitance $C(V)$ beyond the linear response limit of 
Eqs.~(\ref{diff-capacitance}) and (\ref{capacitance}), one takes the Legendre transform  ${\cal E}$ of the energy $E(Q)$ with respect to $Q$ and identifies the capacitance $C(V)=Q(V)/V$ from ${\cal E}(V)= E(Q(V)) - QV$. 

For the sake of clarity, we explicitly note that the value of the differential capacitance for specific values of  $V$, $Q$, or the density of charge carriers, $n$, is given by the ratio $dQ/dV$, where $dQ$ is the  amount of charge that a voltage change $dV$ adds to a plate of the capacitor, which is characterized by $V$, $Q$, or $n$.   We also note  that the voltage $V$ between the two plates, i.e., the difference between the electrochemical potentials of the plates, does in general not equal the difference of the electrical potentials of the electrodes, which defines the electrical voltage $V_{\rm e}$. Whereas the voltage $V$ is the voltage that is measured by standard voltmeters and, for example, is given by the output of standard voltage sources, the electrical voltage $V_{\rm e}$ is the electrical potential difference between the plates, the gradient of which is the electric field between the plates. The electrochemical potential difference consists of the electrical voltage and the difference of the chemical potentials of the electrodes. In standard capacitors, the shifts of the chemical potential upon charging are negligible. However, if the chemical potential difference becomes significant, as is the case when the charging of the capacitors cause a non-negligible shift of the chemical potentials, the measured voltage will differ from the electric potential difference.

\section{Capacitors with Electrodes Comprising Electron Systems with Uniform Background}
\label{sec:e-gas}

Electron gases with uniform, positively charged background (jellium model) serve as model systems which allow to evaluate thermodynamic electronic properties in 2D~\cite{comment2Dphases} and 3D~\cite{comment3Dphases} without the complications arising from the coupling to lattice degrees of freedom. In the following we will calculate the capacitance for 2D or 3D electron gases in the jellium model to make several of our statements on the quantum mechanical properties of capacitors more lucid. The ground state of the electron gases in a configuration with two unbiased but electrostatically coupled sheets, each comprising a dilute electron gas, has been investigated intensely. In particular, it has been laid open that the exchange term $ \sum_i E_{{\rm x},i}$ can drive the double-layer electron gas into ferromagnetic and charge separated states.~\cite{Ruden91,Zheng97,Hanna00} 

A spontaneous spin polarization of a dilute electron gas was already suggested by Felix Bloch~\cite{Bloch29} and has been included in several recent investigations of double layer systems. An exchange driven instability towards a state in which all electrons occupy a single layer, i.e., a state with a maximally charged capacitor has been proposed~\cite{Ruden91} but it has been shown that the ground state in an unbiased double layer system with
$E=E_{\rm Coul} + \sum_i E_{{\rm kin},i} +  \sum_i E_{{\rm x},i}$ is of a different nature:~\cite{Zheng97}
it is a coherent quantum mechanical superposition 
of the two states that each carry the electrons on either of the two layers. This ground state has been introduced and predicted for bilayer quantum Hall systems.~\cite{Halperin94,Moon95} It is characterized by a spontaneous interlayer phase coherence (SILC) of two layers which are each entirely isolated except for their interlayer Coulomb interaction. The mean value of the charge in the  SILC state is zero on each of the two layers. The Coulomb energy $E_{\rm Coul}$ therefore vanishes in this state. 
The possible magnetic, charge separated, and SILC states have been categorized in Hartree-Fock theory at zero magnetic field,  with and without applied bias.~\cite{Hanna00} A finite bias may support charge separated states. 

Spin polarized and SILC states offer exciting possibilities to explore and use quantum capacitances. Nevertheless, in the rest of this article we will focus on unpolarized states without interlayer coherence. While the SILC states will be the true ground states, coherency will be lost in capacities operating with finite electric fields in noisy electronic circuits at room temperature. The states with charged electrodes may be generated through an applied voltage in cases where the differential capacitance $C$ for $Q\rightarrow 0$ is negative (see the following subsections). The states with charged electrodes are expected to 
persist up to high temperatures of the order of
$|E_{\rm Coul} + \sum_i E_{{\rm kin},i} +  \sum_i E_{{\rm x},i} + \sum_i E_{{\rm c},i}|/k_B$. 
In compounds with strongly electronic correlations, the correlations may favor either the SILC states or the states with charged electrodes.  We consider the unpolarized homogeneous electron gas on two nanoscopically close plates as a ``model capacitor'' and refer to possibly relevant, correlated electron systems in the following chapters.

We conceive that practical capacitors with negative, total capacitance can eventually be realized, 
in particular by using strongly correlated systems. The existence of devices with negative 
total capacitance has already been addressed in the literature, concerning electronic circuits (see, e.g, Ref.~\onlinecite{Beavis54}), 
electron injection through interfaces (see, e.g, Ref.~\onlinecite{Omura00}), and structures involving ferroelectrics
(see, e.g, Refs.~\onlinecite{Datta07}, \onlinecite{ Zhirnov08}). Here, we consider a far more conventional system, standard capacitors, such as plate
capacitors comprising conventional dielectric materials and operated in steady state.

\subsection{Electrodes with Two-Dimensional, Homogeneous Electron Systems}
\label{sec:2D}

The capacitance $C$ for a capacitor with  two-dimensional electron systems on each plate is determined  by the
inverse capacitances given in Eq.~(\ref{capacitance}). 
To shed light on the physical origin of the rhs of Eq.~(\ref{capacitance}) and to present a method to approximately calculate these terms, we introduce the thermodynamic compressibility $\varkappa$ of an electron system. As follows from Seitz's theorem~\cite{Mahan},  $\varkappa^{-1}$  is given by the first derivative of the chemical potential $\mu$ or by the second derivative of the total energy $E$, both taken with respect to the electronic density $n=N/A$:
 $\varkappa^{-1}= n^2 {\partial {\mu}}/{\partial n} = n^2\, {d^2 (E/A)}/{d n^2}$, which is a standard thermodynamic relation in the limit of zero temperature. For 2D-systems we rewrite Eq.~(\ref{capacitance}) in terms of the inverse compressibility of a 2D conductor:\cite{Stern72}
\begin{equation}
\label{cap-2D}
A/C^{{\rm (2D)}} - AC^{-1}_{\rm geom} \,=\,   \sum_i \frac{\varkappa_i^{-1}}{(e n_i)^2} 
\end{equation}

The inverse kinetic and exchange capacitances can be determined from Eq.~(\ref{cap-2D}), using the well-known expressions for the energy functional of a 2D homogeneous electron system (see, e.g., Ref.~\onlinecite{Tanatar89}):
\begin{eqnarray}
4\pi \epsilon_0 A/C_{{\rm geom}}  & = & \frac{4\pi d}{ \epsilon_r}
\label{cap-geom}\\
4\pi \epsilon_0 A/C_{{\rm kin},i}^{{\rm (2D)}}  & = &  \frac{\pi a_B}{m_i^\star/m} =  \frac{4\pi \epsilon_0}{e^2} \frac{1}{ \rho_i^{(2D)}(\varepsilon_F)}
 \label{cap-2D-kin}\\
4\pi \epsilon_0 A/C_{{\rm x},i}^{{\rm (2D)}}  & = &  - \bigl({\frac{2}{\pi}}\bigr)^{\frac{1}{2}} \, \frac{1} {\sqrt{n_i}} \,\frac{1}{\epsilon_{{\rm eff},i}}
\end{eqnarray}
where $a_B=4\pi\epsilon_0\hbar^2/(me^2)\simeq 0.5$\AA\ is the bare Bohr radius, $m_i^\star$ is the effective electronic mass and $\epsilon_{{\rm eff},i}$ the effective dielectric constant in plate $i$. This constant characterizes the screening in the metallic 2D plate that does not arise from the screening by the conduction band electrons themselves. The band electron ``self-screening'' is included in the correlation term $C_{\rm c}$ discussed below. It is notoriously difficult to estimate $\epsilon_{{\rm eff},i}$. For very dilute 2D electron systems in GaAs/AlGaAs heterostructures that may have nominal densities of $10^{11}$~cm$^{-2}$, $\epsilon_{{\rm eff}}$ is given approximately by the dielectric constant of the AlGaAs layer.~\cite{Eisenstein92} The effective band mass $m_i^\star$ can be more easily determined. Conveniently it is chosen such that the resulting DOS matches the DOS $\rho_i^{(2D)}(\varepsilon_F)$ of the conductor. A summation over bands is used in the DOS in case several conduction bands are present or a band degeneracy exists.~\cite{comment2} Such a band (degeneracy) factor decreases the inverse kinetic capacitance and, consequently, increases the total capacitance $C$. 

The kinetic capacitance $C_{{\rm kin},i}^{{\rm (2D)}}$, which has been introduced by Luryi~\cite{Luryi88} as ``quantum capacitance'', is invariably positive.
The exchange capacitance  $A/C_{{\rm x}}$, however, is always negative, irrespective of the dimensionality of the electron gas. The exchange capacitance varies inversely with the square root of the electronic density. Correspondingly, based on Eq.~(\ref{capacitance}) we recognize that capacitors with negative capacitances can be built. The negative capacitance can be obtained, for example, by fabricating capacitors with electrode materials that have a very low density of states. A negative $C$ implies that the plates of the capacitor self-charge once they are connected. In these systems, the inverse exchange $A/C_{{\rm x}}$ plus correlation $A/C_{\rm c}$ capacitances dominate the inverse kinetic capacitances $A/C_{{\rm kin}}$ plus  the geometrical capacitance, so that the total capacitance $C$ becomes negative. 

In a homogeneous electron gas, electron-phonon contributions are not present by definition. Correspondingly, 
$A/C_{{\rm el-ncc}}$ in Eq.~(\ref{capacitance}) is disregarded here.

The last term to be considered in $A/C$ of Eq.~(\ref{capacitance}) is the correlation term. If present, $C_{\rm c}$ is particularly important, and obviously even more so in those cases in which $A/C_{{\rm x}}$ and $A/C_{{\rm kin}}$ almost cancel each other.

\begin{figure}[t]
\centering
\includegraphics[width=0.99\columnwidth]{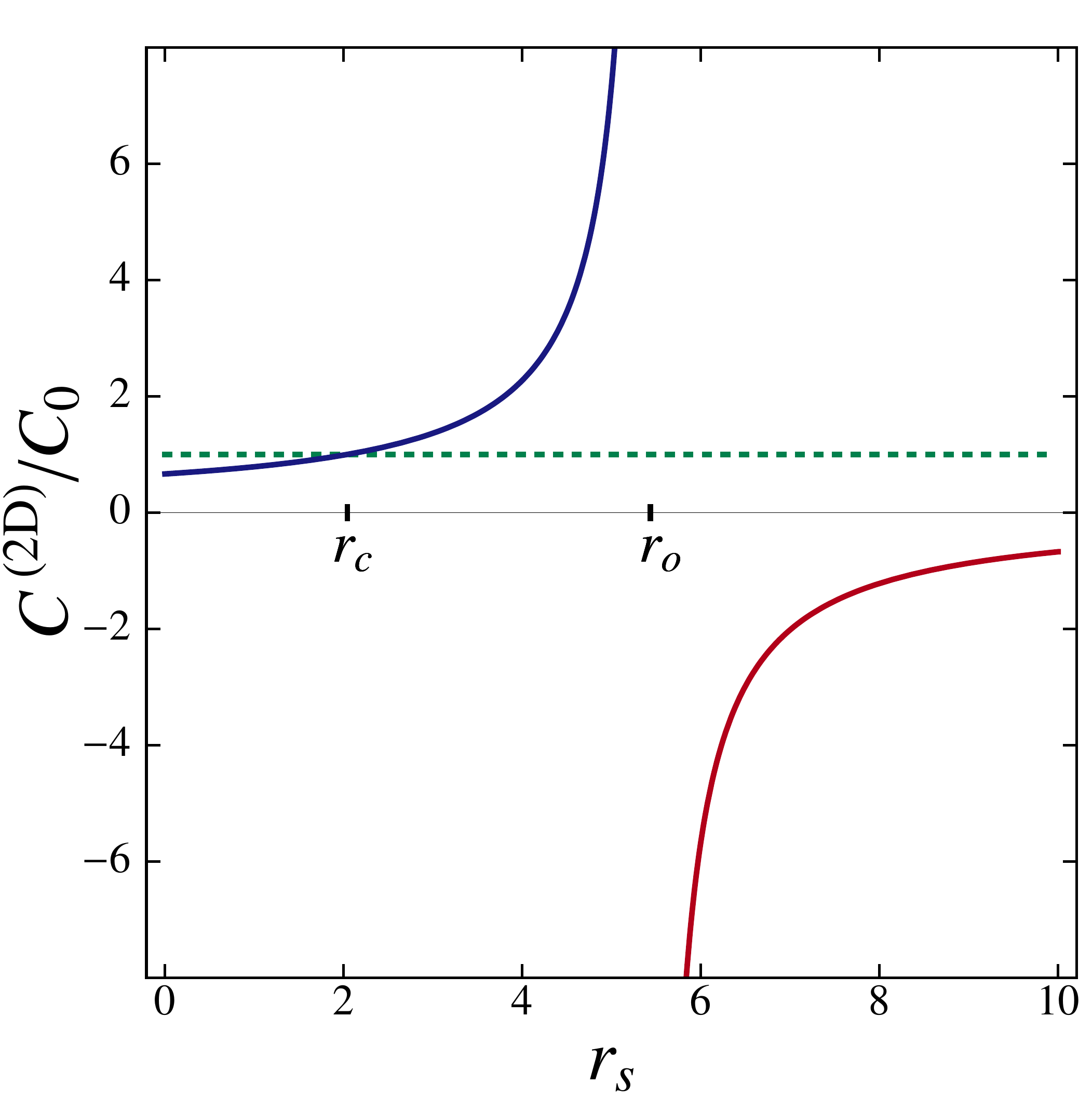}
\caption
{Dependence of the capacitance of an ideal two-dimensional capacitor (a parallel plate capacitor with two 2D-electrodes) on the carrier spacing. The capacitance is a function of the electron density on the electrodes, parameterized by the dimensionless interparticle distance $r_s=\bigl({1}/{a_B^2 \pi n}  \bigr)^{\frac{1}{2}}$ in bare atomic units. In the calculation presented, the effective mass is
taken to be $m^\star/m=1$, the effective dielectric constant in the electrodes to be 
$\epsilon_{{\rm eff},i}=1$, and the effective distance to be $d^\star=a_B$. The dotted line shows the capacitance of the conventional, classical capacitor $C_0 \equiv C_{\rm geom} $.}
\label{Fig2Dcap}
\end{figure}

To estimate the correlation capacitance of the 2D homogeneous electron gas, the parameter $r_s$ is introduced. It is a dimensionless number that characterizes the interparticle distance:
\begin{equation}
\label{rs2D}
r_s[n_i] = \bigl(\frac{1}{a_B^2 \pi n_i}  \bigr)^{\frac{1}{2}}
\end{equation}
Tanatar and Ceperley~\cite{Tanatar89} have cast the  $r_s$-dependence  of the correlation energy into an approximate functional form  which fits their Monte Carlo data. With the energy functional of Ref.~\onlinecite{Tanatar89}, the inverse capacity is identified by using the correlation contribution to Eq.~(\ref{cap-2D}). The exchange and correlation capacitances are found to equal
\begin{equation}
4\pi\epsilon_0 A/C_{{\rm x},i}^{{\rm (2D)}} = -\sqrt 2 \; a_B\, r_s/{\epsilon_{{\rm eff},i}}
 \label{cap-2D-xc}
\end{equation}
and
\begin{eqnarray}
\label{cap-2D-corr}
&&\!\!\!\!\!\!\!  4\pi\epsilon_0 A/C_{{\rm c},i}^{{\rm (2D)}} =\\
&& \frac{\pi  a_B r_{s}^2}{\epsilon_{{\rm eff},i}}\;  \frac{a_0}{y} 
\Bigl(-\frac{3}{32} u +(1+\frac{3}{4}u)\frac{x}{y} +(1+u)(\frac{x^2}{y^2}-\frac{z}{y})\Bigr),
       \nonumber
\end{eqnarray}
respectively.
Here, the polynomial functions $u,x,y,z$ are defined as:
\begin{subequations}
\label{uxyz}
\begin{alignat}{2}
u[r_s] &= a_1 r_s^{\frac{1}{2}}& \\
y[r_s] &= 1 + a_1 r_s^{\frac{1}{2}} + a_2 r_s + a_3 r_s^{\frac{3}{2}}& \\
x[r_s] &= \frac{1}{4} a_1 r_s^{\frac{1}{2}} + \frac{1}{2} a_2 r_s + \frac{3}{4} a_3 r_s^{\frac{3}{2}}& \\
z[r_s] &= \frac{5}{32} a_1 r_s^{\frac{1}{2}} + \frac{3}{8} a_2 r_s + \frac{21}{32} a_3 r_s^{\frac{3}{2}}  &
\end{alignat}
\end{subequations}
and the coefficients $a_l$ are~\cite{Tanatar89}
\begin{equation}
\label{a123}
a_0=-0.3568,\;\;  a_1=1.1300, \;\; a_2=0.9052, \;\; a_3=0.4165
 \nonumber\end{equation}
The inverse capacity of each of the plates contributes a term $A/C_{{\rm c},i}^{{\rm (2D)}}$ to the total inverse capacity. 
For large $r_s$, i.e., for dilute, homogeneous electron systems, $A/C_{{\rm c}}^{{\rm (2D)}}\propto -r_s \propto -1/\sqrt{n}$. This implies that in systems with very few carriers, $C$ is driven by $C_{\rm x}$ and $C_{\rm c}$ to small, negative values (Fig.~\ref{Fig2Dcap}).

Eqs.~(\ref{cap-2D}) and (\ref{cap-2D-kin}), (\ref{rs2D}) and  (\ref{cap-2D-corr}) reveal that the total capacitance can differ considerably from the geometrical capacitance. It is only for the case that ${C_{\rm geom}}/{A}$  is much smaller than 
$ |\varkappa_i | \,  (e n_i)^2$
that the total capacitance is bound to be given by the geometrical capacitance. 
The strong dependence of $C$ on $r_s$ is illustrated in Fig.~\ref{Fig2Dcap}  for a symmetric capacitor. 
As expected from the previous discussion, with increasing $r_s$ 
the capacitance diverges and becomes negative when the negative terms $2(A/C_{{\rm x}}+A/C_{{\rm c}})$ compensate the positive terms $(2A/C_{{\rm kin}}+A/C_{{\rm geom}})$ at  $r_s =r_0$.
The $r_s$ value at which the capacitance changes sign is higher than the one at which the compressibility becomes negative. We assign to the latter the dimensionless length scale $r_c$, and $r_0> r_c$.
In fact, the transition into a state of negative electronic compressibility has already been experimentally observed 
at interface electron gases in Si-MOSFETs and III--V 
heterostructures~\cite{Kravchenko89,Eisenstein92,Shapira96}. The observation of a negative $\varkappa$ does not necessarily contradict the thermodynamic stability criterium of a positive total compressibility. The latter condition applies to the sum of the compressibilities of all electronic and ionic subsystems. Even in thermodynamic equilibrium, the compressibility  $\varkappa$ of the electronic subsystem on one or on both of the electrodes may be negative~\cite{comment2Dphases} if it is compensated by other contributions to the total compressibility.

It has to be expected that the electronic states on the plates change qualitatively at the transition into the regime of negative electronic compressibility~\cite{Bello81} at $r_c$. The new ground state is characterized by exponentially damped charge density modulations  (cf.~[\onlinecite{Schakel01}]). A further transition into a self-charged state takes place at $r_0$ where the capacity becomes negative.

As shown by Fig.~\ref{Fig2Dcap}, the capacitance exceeds the classical capacitance ($C >C_{{\rm geom}}$)
for $r_c< r_s <r_0$. 
This result relies on the assumption that, also at these carrier densities, the electron system is homogeneous.
For non-homogeneous electron systems, the phase transition into an electronic state with negative compressibility has not been investigated rigorously. 
It cannot be excluded that for $r_s > r_c$ the capacitance behaves differently 
than shown by Fig.~\ref{Fig2Dcap}. 
In particular, a metal-insulator transition~\cite{Kravchenko94} with a spatially inhomogeneous insulating phase~\cite{Ilani00} has been observed which is probably controlled by strong disorder. 
Also, a phase transition into a charge density wave state seems possible.~\cite{Neilson} 
The functional properties of the capacitance nevertheless reflect the energy dependence of the electronic state on the charge carrier density of the electrodes. 

\begin{figure}[t]
\centering
\hbox{\hskip-4.0pt\includegraphics[width=1.05\columnwidth]{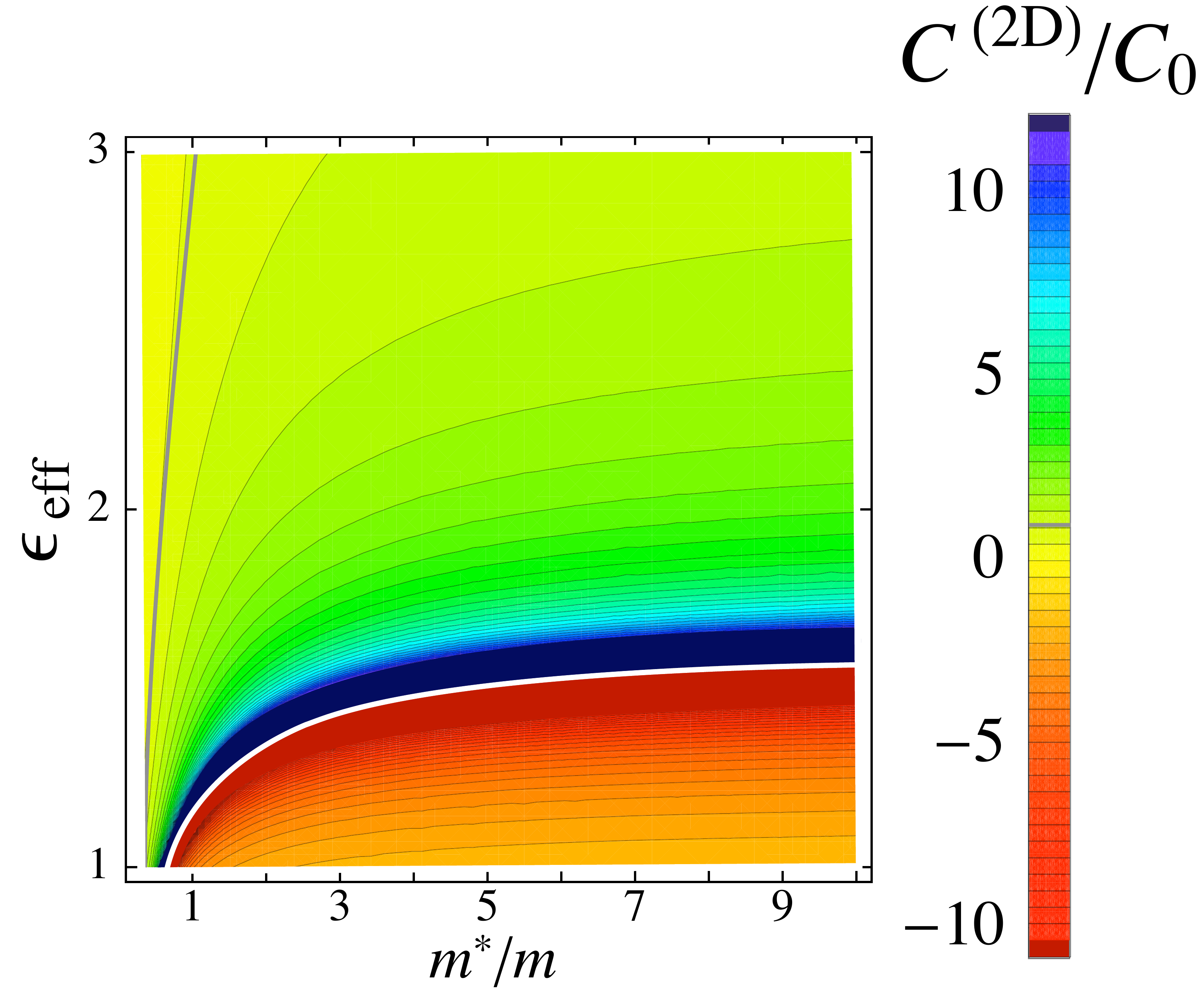}}
\caption
{
Contour plot, showing for an ideal two-dimensional symmetric capacitor the lines of constant
capacitance. The capacitance $C^{\rm 2D}$ is a function 
of the effective mass $m^\star$ and of the effective dielectric constant $\epsilon_{{\rm eff}}$ of the electrode materials (assumed to be equal on both electrodes).
Here, the dimensionless interparticle distance $r_s=\bigl({1}/{a_B^2 \pi n}  \bigr)^{\frac{1}{2}}$ in bare atomic units is
taken to be $r_s=7$  and the effective distance of the capacitor plates to be $d^\star\equiv d/\epsilon_r=1.23\,a_B$, 
where $\epsilon_r$ is the dielectric constant of the dielectric material between the plates. 
The white line traces the values of  $m^\star/m$
versus $\epsilon_{{\rm eff}}$ at which the capacitance diverges (cf.~Fig.~\ref{Fig2Dcap}).
}
\label{2Dcontour}
\end{figure}

For the ideal capacitor with two-dimensional electrodes, the 
kinetic term has a negligible influence on the capacitance for large density of states or, equivalently for large effective mass (see Fig.~\ref{2Dcontour}). This applies for $\rho^{(2D)}(\varepsilon_F)\gg 2\epsilon_0\epsilon_r/(e^2 d)$, i.e., for $m^\star/m \gg \epsilon_r a_B/(2 d)$.

Leaving the model capacitor, the description of a realistic quasi-2D electronic system has  to consider also the finite thickness of the quantum well which hosts the 2D electron system. As discussed, the Hartree term contains such corrections which are  included in $d^\star$. Also the exchange and correlation terms depend on the finite thickness which may be considered by a ``form factor'' that acts as a prefactor
for the respective energies. Such a form factor has been introduced by Stern~\cite{Stern72} for the exchange term and has been applied by Eisenstein {\it et~al.\/}~\cite{Eisenstein92} in the evaluation of  their experiments. This finite
thickness correction does not alter in a qualitative way the dependence of the capacitances as given by Eq.~\ref{cap-2D-xc}. 

Furthermore, also the influence of the charge carrier density on the band energy (band bending) has not been considered in the model capacitor model. This effect depends on the details of the electronic structure of the electrode material. A detailed analysis is given in Ref.~\onlinecite{Eisenstein92}.

\subsection{Electrodes with Three-Dimensional, Metallic Electron Systems}
\label{sec:3D}

Quantum corrections to the capacitance that result from the kinetic energy 
of the additional carriers of a charged 3D capacitor,
have been investigated in Refs.~\onlinecite{Stern72,Hauser98,Buettiker93}.
If the corresponding states  are confined to the surface, i.e., 
if they form subbands of surface bound states, they are two-dimensional states. 
Their existence requires that their energies are 
well separated on the temperature scale $k_B T$ and that the inelastic scattering 
between states of these subbands is negligible.
In that case the evaluation of Sec.~\ref{sec:2D} applies. In this subsection we consider the case that the electronic states are not bound to the surface of the electrode.

Also in 3D systems, the geometric capacitance is given  by Eq.~(\ref{cap-geom}), 
\begin{equation}
\label{Cgeom}
4\pi\epsilon_0 A/C_{\rm geom}=4\pi d^\star
\end{equation}
except that the charge does not completely reside  in the surface planes of the electrodes. A simple 
approach for 3D systems assumes the charge density to decay into the bulk with a length scale of the order of the screening length $\lambda$. The effective distance between the capacitor plates is thereby slightly enhanced by $2\lambda$, because the static Coulomb energy is determined by the charge distribution which is influenced by the boundary condition of the electron wave function at the 
surface.~\cite{Ku64,Tsong69,Stern72,Krisch96,Hauser98,Black99}  

The results of such an approach have to be contrasted with those of Lang and Kohn.~\cite{Lang73} They include the formation of a surface dipole barrier that is caused by the ``spilling out'' of the mobile electrons into the vacuum. Already in the uncharged state this dipole layer forms a non-homogenous charge profile. 
This charge profile modifies the distribution of the induced screening charge for finite voltage so that the effective distance between the electrodes is reduced rather than enhanced by $2\lambda$, assuming that the difference between the edge of the uniform background charges defines the distance $d^\star$. Whereas the width of the screening charge density scales approximately with the Thomas-Fermi screening legnth~\cite{Liebsch87}---apart from the characteristic Friedel oscillations in the interior of the metal---the center of mass position of the induced charge is not easy to determine. Its calculation requires a fully self-consistent treatment that includes the exchange and correlation terms.

To elucidate the behavior of capacitors with metallic electrodes that contain 3D electron gases,~\cite{comment3Dphases} we consider in the following a model capacitor. In this capacitor, the 3D bulk compressibility of the electronic systems is taken to be constant up to the electrode surfaces; consequently the formation of a surface dipole layer is not included. With this approximation, the precise distance between the mirror planes of the electrodes cannot be  identified, as in Ref.~\onlinecite{Lang73}  for the jellium-model electrodes separated by vacuum. However a 
comparison of the qualitative dependence of the effective distance on exchange and correlation effects is of considerable interest and will be presented below. This allows to introduce an effective model which captures an important part of the physics and is much easier to treat than the fully self-consistent evaluation of 
Ref.~\onlinecite{Lang73}. It may seem that the approach, which is presented below and builds on the evaluation of the density-density correlation, cannot account for the displacement of the center of mass of the induced charge density as introduced in Ref.~\onlinecite{Lang73}. The density-density correlation in the uncharged state of the electrodes is, however, related via the fluctuation-dissipation theorem to the displacement of the induced charge for finite, positive compressibility. 

In three dimensions, the parameter $r_s$ is defined by
\begin{equation}
\label{rs3D}
r_s[n_i] = \bigl(\frac{3}{4\pi n_i a_B^3}  \bigr)^{\frac{1}{3}}
\end{equation}
and presents, like in the 2D systems, the dimensionless length scale of the carrier spacing. We assume that the electrodes in the uncharged state consist of a homogeneous electron system with a uniform positive background which compensates the electron charge (jellium model). Correspondingly, $r_s$ and the energy $E$ are not functionals of $n_i[x]$ but just functions of the average electronic densities $n_i$ as in Sec.~\ref{sec:2D}. In a microscopic evaluation of the work function of the electrodes, this approximation would
miss the electrostatic potential across the metal surface, because the homogeneous electron system 
(up to the edge of the positive background) does not include the surface dipole barrier. 
However, these terms should be
included phenomenologically in the difference of the work functions for the two electrodes in the capacitance. For electrode plates of the same material and the same crystallographic orientation of their inside parallel faces, these terms cancel.


With the 3D electronic density $n=N/{\cal V}$, the compressibility relation is: 
$\varkappa^{-1}= n^2 {\partial {\mu}}/{\partial n} = n^2\, {d^2 (E/ {\cal V})}/{d n^2}$.
For a capacitor with 3D plates, the screening of the charge causes the charge density 
to be non-uniform  perpendicular to the plates on the scale of the screening lengths $\lambda_i$.
With the above introduced assumption of constant compressibility, the relation between the capacitances and compressibilities has the approximate form

\begin{equation}
\label{cap-3D}
A/C^{{\rm (3D)}} - AC^{-1}_{\rm geom} \,=\,    \sum_i \frac{1}{e^2 \lambda_i} \frac{\partial\mu_i}{\partial n_i}
 =\,    \sum_i \frac{\varkappa_i^{-1}}{\lambda_i (e n_i)^2} 
\end{equation}
In case the density of states (DOS) $\rho^{(3D)}(\varepsilon_F)$ (the DOS at the Fermi energy $\varepsilon_F$ for both spin directions) is spatially uniform and if the screening follows the Thomas-Fermi screening of a free electron gas with ${\lambda_i^{\rm (TF)}}^{-2}= \epsilon_0^{-1} e^2 \rho_i^{(3D)}(\varepsilon_F)$, Eq.~(\ref{cap-3D}) 
is equivalent to the relation $A/C^{{\rm (3D)}} - AC^{-1}_{\rm geom} \,=\,   \epsilon_0^{-1} \sum_i \lambda_i^{\rm (TF)} $ given by B{\"uttiker}~\cite{Buettiker93}. 
The relevant surface volume of the electrode $i$ is given by the surface area times the  screening length $\lambda_i$.

Eq.~(\ref{cap-3D}) yields the value of the inverse capacitance.
The kinetic term in the energy functional generates for each electrode a capacitance, the inverse value of which is:
\begin{equation}
\label{C3Dkin} 
4\pi \epsilon_0 A/C_{{\rm kin},i}^{{\rm (3D)}}  = 
\Bigl(\frac{\pi^4}{3}\Bigr)^{\frac{1}{3}}\;\frac{a_B}{\lambda_i}\frac{n_i^ {-\frac{1}{3}}}{m_i^\star/m} =  
\frac{4\pi\epsilon_0}{e^2\lambda_j} \frac{1}{ \rho_i^{(3D)}(\varepsilon_F)}
\end{equation}
A straightforward conversion to the Thomas-Fermi approach, which also accounts for the polarizability of the underlying lattice and of the ionic cores through a corresponding effective dielectric constant $\epsilon_{{\rm eff},i}$ (see, e.g., Ref.~\onlinecite{Black99}), identifies:
\begin{equation}
\label{lambda}
\lambda_i=\epsilon_{{\rm eff},i}^{\frac{1}{2}}\; \lambda_i^{{\rm (TF)}}
\end{equation}
where the bulk Thomas Fermi screening length is
\begin{equation}
\label{lambdaTF}
\lambda_i^{{\rm (TF)}}=\frac{1}{\sqrt{e^2  \rho_i^{(3D)}(\varepsilon_F)/\epsilon_0}  }
= \frac{1}{2}\, \bigl(\frac{2\pi}{3}\bigr)^{\frac{1}{3}} \sqrt{\frac{m}{m_i^\star}}\, a_B \sqrt{r_s}
\end{equation}
Eq.~(\ref{C3Dkin}), jointly with Eqs.~(\ref{lambdaTF}) and (\ref{lambda}), yields the simple relation:~\cite{Black99} 
\begin{eqnarray}
\label{C3DkinTF} 4\pi\epsilon_0 A/C_{{\rm kin},i}^{{\rm (3D)}}  &= &
\frac{4\pi \lambda_i^{\rm (TF)}}{\sqrt{\epsilon_{{\rm eff},i}}}
\end{eqnarray}
Here we have used the density of states (DOS) relation for an electron gas: 
\begin{equation}
\label{DOS}
\rho_i^{(3D)}(\varepsilon_F)=\bigl(\frac{3}{\pi^4}\bigr)^{\frac{1}{3}} \, 
                      \frac{4\pi\epsilon_0 n^{\frac{1}{3}}}{e^2 a_B} \frac{m_i^\star}{m}
=\frac{1}{\pi}\, \bigl(\frac{3}{2\pi}\bigr)^{\frac{2}{3}} \frac{4\pi\epsilon_0}{e^2 a_B^2\, r_s}  \frac{m_i^\star}{m}
\end{equation}
We derived the relation Eq.~(\ref{C3DkinTF}) from the (constant) compressibility of the kinetic energy term, whereas the standard derivation determines first the induced charge density in Thomas-Fermi approximation.
Both schemes are obviously equivalent if the screening length is identified as $\lambda_i^{{\rm (TF)}}$.

The quantum kinetic term (Eq.~(\ref{C3Dkin})) depends on the DOS and always lowers the capacitance as compared to its classical value. As shown by Eqs.~(\ref{C3Dkin}) and (\ref{lambdaTF}), multiple or
degenerate conduction bands in the DOS increase $C^{\rm (3D)}$.
Conversely, small effective masses $m^\star$ of the carriers reduce $C^{\rm (3D)}$ (cf.~Fig.~\ref{3Dcontour}).

\begin{figure}[t]
\centering
\includegraphics[width=0.97\columnwidth]{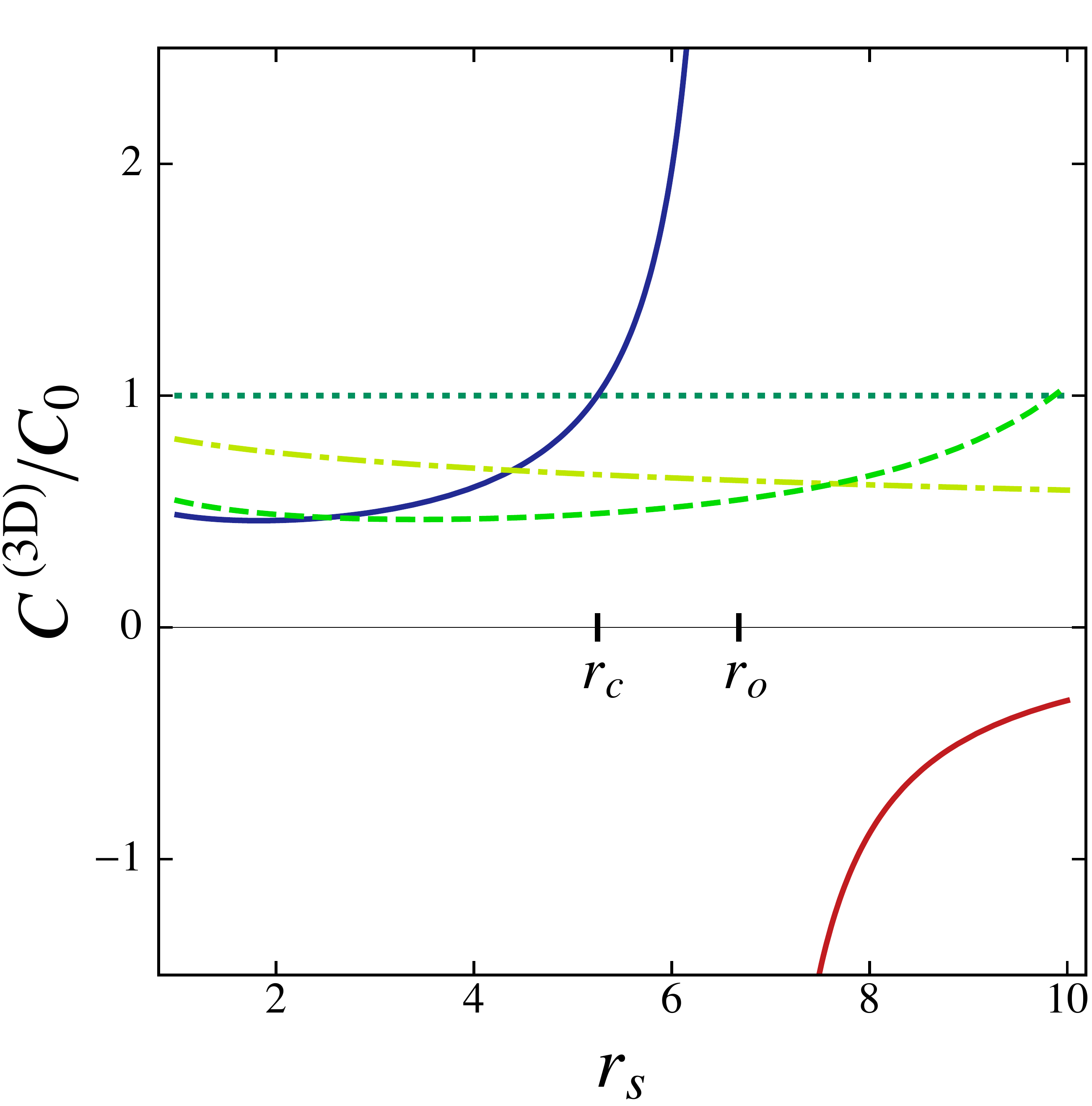}
\caption
{Dependence of the capacitance of a three-dimensional, parallel plate capacitor on the carrier spacing, according to the serial connection of the capacitances of Eqs.~(\ref{Cgeom}), (\ref{C3Dkin}), (\ref{C3Dxc}) and (\ref{C3Dcorr}). The capacitance is a function of the electron density on the surface of the electrodes, parameterized by the dimensionless interparticle distance $r_s=\bigl({3}/{4\pi n a_B^3}  \bigr)^{\frac{1}{3}}$. In the calculation shown, 
the effective mass is $m_i^\star/m=1$, the effective dielectric constant in the plates is 
$\epsilon_{{\rm eff},i}=1$, and their effective distance is $d^\star=a_B$. 
The dotted line displays the capacitance of the corresponding classical 
capacitor $C_0\equiv C_{\rm geom}$. The dashed line refers to a capacitor with $\epsilon_{{\rm eff},i}=2$,
and the dot-dashed line to a capacitor with $\epsilon_{{\rm eff},i}=30$.}
\label{Fig3Dcap}
\end{figure}

\begin{figure}[t]
\centering
\includegraphics[width=0.99\columnwidth]{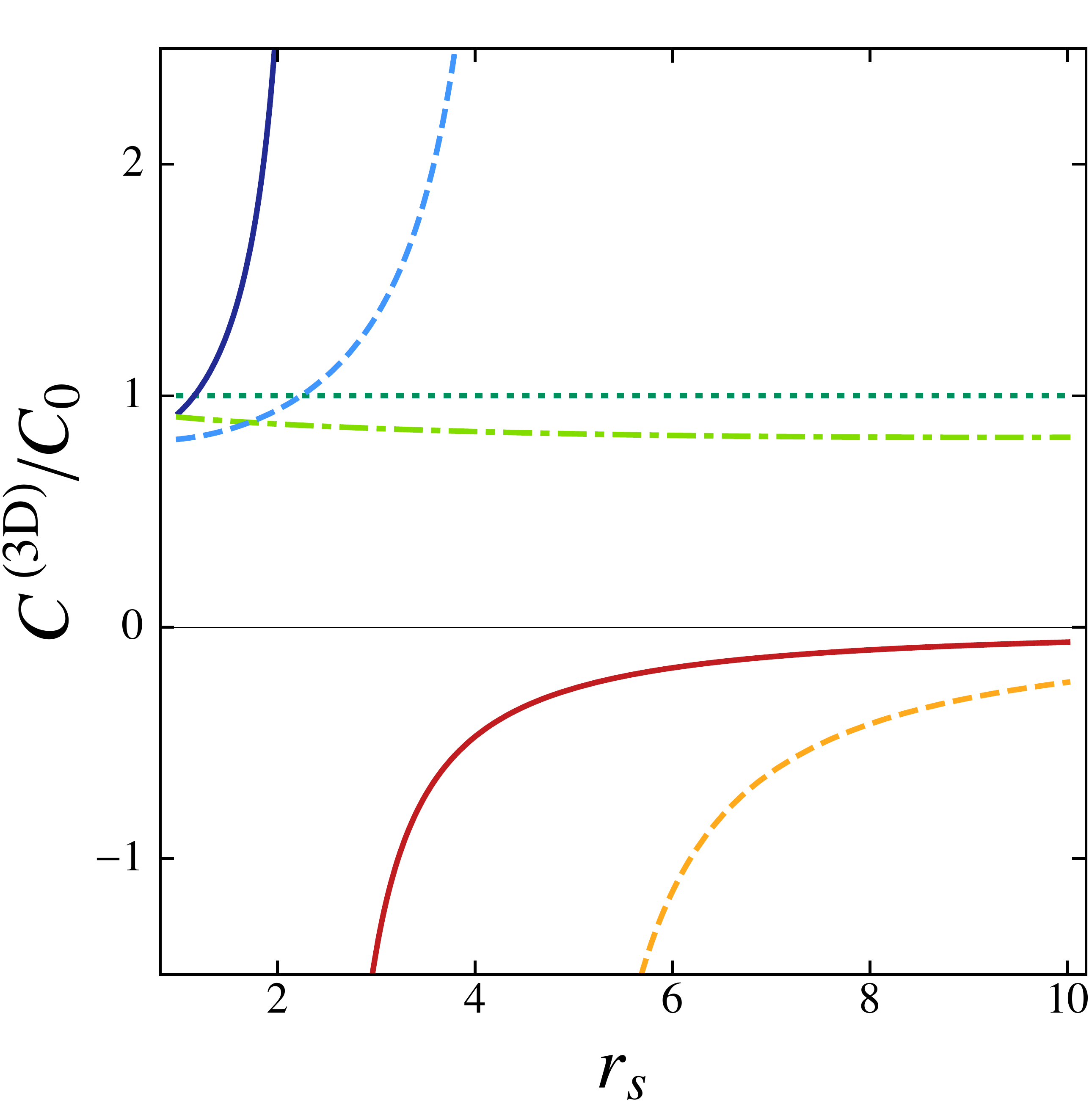}
\caption
{Dependence of the capacitance of a three-dimensional, parallel-plate capacitor as a function of carrier spacing. The effective distance between the plates used in the calculation is $d^\star=a_B$ and $m_i^\star/m=5$.
The effective dielectric constants are equal for both electrodes with $\epsilon_{{\rm eff},i}=1,2$, and $30$ for the solid, dashed, and dot-dashed lines, respectively. The dotted line shows the capacitance of the corresponding classical capacitor 
$C_0\equiv C_{\rm geom}$.}
\label{Fig3DcapHighMass}
\end{figure}

Coulomb interactions beyond the classical approximation qualitatively alter the properties of the capacitance, because exchange and correlation terms introduce in Eq.~(\ref{capacitance}) negative inverse capacitances.  With increasing electronic correlations the capacitance therefore grows to jump at $r_s > r_0$ to negative values (see Fig.~\ref{Fig3Dcap}).  

The exchange contribution to the capacitance results
from  the standard exchange energy functional~\cite{Mahan} (using Eq.~(\ref{cap-3D})):
\begin{eqnarray}
\label{C3Dxc} 4\pi\epsilon_0 A/C_{{\rm x},i}^{{\rm (3D)}}  &= &
-\frac{1}{(9\pi)^{\frac{1}{3}}}\;\frac{1}{\lambda_i\epsilon_{{\rm eff},i}\,n_i^{\frac{2}{3}}}\nonumber\\
&=& 
-\Bigl(\frac{16\pi}{81}\Bigr)^{\frac{1}{3}} r_s^2\;\frac{a_B^2}{\lambda_i\epsilon_{{\rm eff},i}}
\end{eqnarray}

\begin{figure}[t]
\centering
\hbox{\hskip-4.0pt\includegraphics[width=1.05\columnwidth]{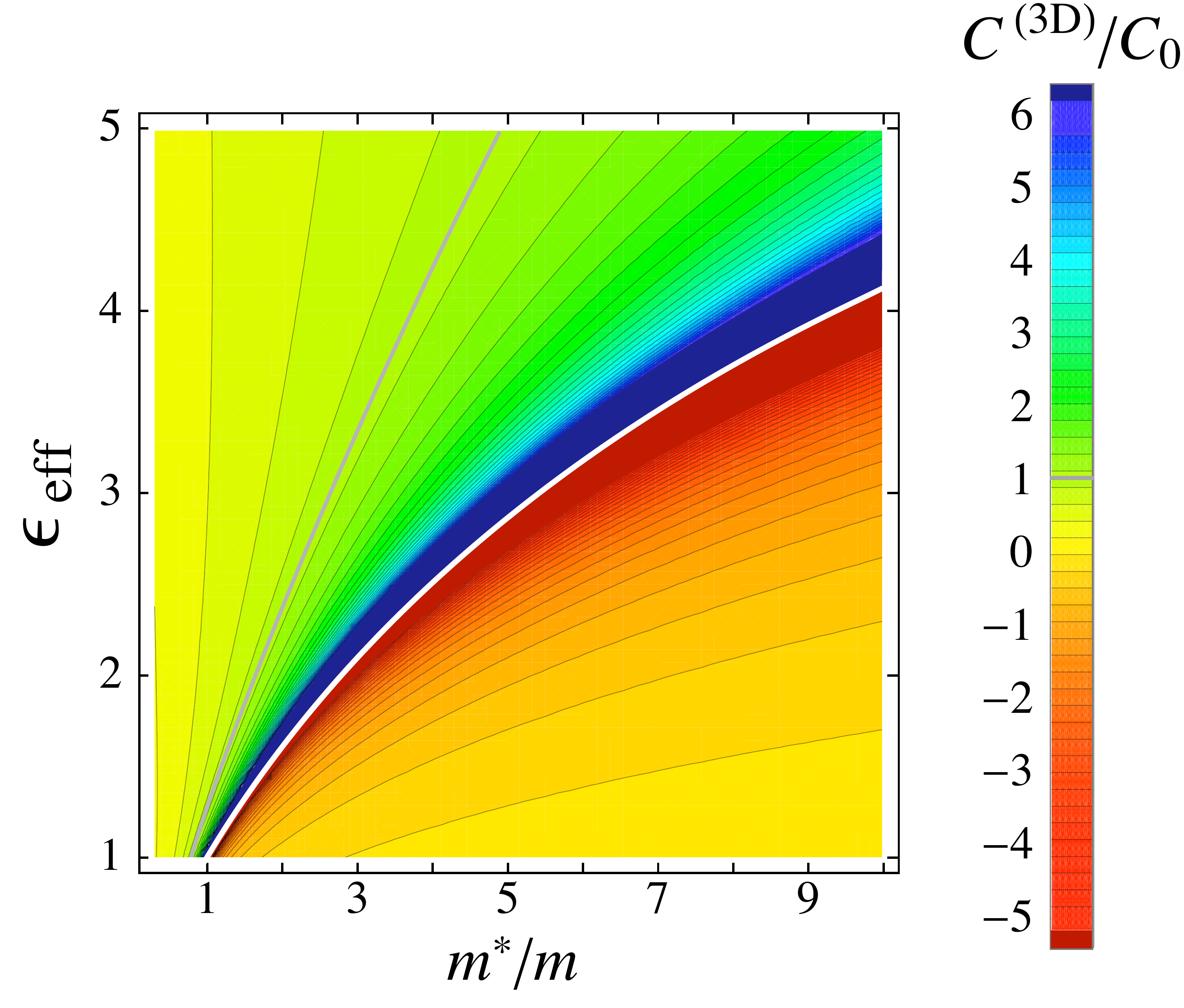}}
\caption
{
Contour lines of equal capacitance  for the three-dimensional model capacitor. The capacitance $C^{\rm 3D}$ is a function 
of the effective mass $m^\star$ and of the effective dielectric constant $\epsilon_{{\rm eff}}$ of the electrode materials.
Here, the dimensionless interparticle distance 
$r_s=\bigl({3}/{4\pi n a_B^3}  \bigr)^{\frac{1}{3}}$
in bare atomic units is
taken to be $r_s=7$  and the effective distance of the capacitor plates to be $d^\star\equiv d/\epsilon_r=1.23\,a_B$.
The gray line traces the values of  $m^\star/m$
versus $\epsilon_{{\rm eff}}$ at which the capacitance diverges.
}
\label{3Dcontour}
\end{figure}

The correlation contribution is derived from the approximate functional form of the 3D energy 
fit to Monte Carlo calculations as given by Ceperley:~\cite{Ceperley78}
\begin{equation}
\label{C3Dcorr}
4\pi\epsilon_0 A/C_{{\rm c},i}^{{\rm (3D)}} =
 \frac{4\pi  a_B^2 r_{s}^3}{3\lambda_i\epsilon_{{\rm eff},i}}\;  \frac{h_4}{f[r_s]^2} 
\Bigl(g[r_s]-s[r_s]+\frac{g[r_s]^2}{f[r_s]}\Bigr)
\end{equation}
The polynomials $f,g,s$ are hereby defined by:
\begin{subequations}
\label{fgs}
\begin{alignat}{2}
f[r_s] &= 1 + \beta_1 r_s^{\frac{1}{2}} + \beta_2 r_s &\\
g[r_s] &= \frac{1}{6}\, r_s^{\frac{1}{2}} + \frac{1}{3}\, \beta_2 r_s & \\
s[r_s] &=\frac{7}{72}\, r_s^{\frac{1}{2}} + \frac{2}{9}\, \beta_2 r_s & 
\end{alignat}
\end{subequations}
and the coefficients $\beta_l $ and $h_4$ are taken from Ref.~\onlinecite{Ceperley78} (where $h_4\equiv\alpha_4 - h_1\beta_2$~\cite{comment3}
as in Table~IV of Ref.~\onlinecite{Ceperley78}):
\begin{equation}
\label{betah}
\beta_1=1.15813,\quad  \beta_2=0.34455, \quad h_4=-0.2942
 \end{equation}
The dependence of the total capacitance $C^{\rm (3D)}$ of a symmetric capacitor on the electron density parameter $r_s$, on the effective mass, and on the dielectric constant of the electrodes is presented in Figs.~\ref{Fig3Dcap} and \ref{Fig3DcapHighMass}, for parameter sets with $m^\star/m=1$ and 5, respectively. 
The contour lines of constant capacitance $C^{\rm (3D)}$ in 
Fig.~\ref{3Dcontour} are qualitatively different from those of the 2D capacitance (Fig.~\ref{2Dcontour}), because the Thomas Fermi screening length with its functional dependence on $m^\star$ and $r_s$ enters the expression for $C^{\rm (3D)}$. Consequently, the transition to a negative capacitance occurs at higher dielectric constants  $\epsilon_{{\rm eff}}$ in the 3D electrodes than for the ideal 2D capacitor (cf.\ the scales of $\epsilon_{{\rm eff}}$ in Fig.~\ref{2Dcontour} and Fig.~\ref{3Dcontour}). For $\epsilon_{{\rm eff}}=1$, however, the transition of a capacitor with 3D electrodes takes place at higher effective mass and larger carrier spacing $r_s$ (i.e.\ lower density) than for the 2D electrodes.

We finally address the question, how these results compare to those of Ref.~\onlinecite{Lang73} where  $\epsilon_{{\rm eff}}=1$ and $\epsilon_{r}=1$ have to be chosen.~\cite{comment1}  Since we did not include the surface dipole layers, we cannot expect to find agreement concerning the effective capacitance length as defined by the inverse total capacitance. We therefore compare the dependencies of the kinetic term and of the combined exchange and correlation terms on $r_s$.
We take the centre of mass position of the induced charge density from a recent publication 
(Table~1 in Ref.~\onlinecite{Schreier87}) in which the values with and without exchange and correlation are explicitly listed. 
The total values of the centre of mass position agree with those of Ref.~\onlinecite{Lang73} at  $r_s =2$ and 4, except for a factor of 2 in Ref.~\onlinecite{Lang73} which has been suppressed in all subsequent publications:

The inverse kinetic capacitance, which they obtain, increases similarly with $r_s$ as the one we are finding.
The inverse capacitance associated with exchange and correlation energies is also negative in their data. 
The magnitude of the capacitance also increase with $r_s$, however at a smaller rate. Our calculation based on a homogenous electron gas apparently overestimates the exchange effect.  The overall trend of the contributions to the capacitance reported in Ref.~\onlinecite{Schreier87} agrees well with our findings, which seems surprising if the crude treatment of the surface inhomogeneity is considered.~\cite{comment4}

\section{Capacitors Comprising Electrodes with Correlated Inhomogeneous Electron Systems}

In a homogeneous electron gas, which resides in a continuous medium, the compressibility of the interacting electrons is dominated by the kinetic and by the exchange contributions. In contrast, an electronic system that resides on a lattice may behave in a rather different manner, as epitomized by the Hubbard model in which the correlation term dominates the physics close to the Mott transition. Inter alia, it is the strong on-site interaction of charge carriers in the lattice models which on the one hand makes these models difficult to treat but on the other hand generates highly intriguing properties. Such strongly correlated systems may display a multitude of phase transitions, sometimes induced by small variations of the electronic density. But also, by altering temperature, magnetic field, pressure, or epitaxial strain, phase transitions may be induced. Intermediate to strong electronic correlations have been identified in systems such as the manganites, which display a colossal magnetoresistance effect, or such as the cuprates with their enigmatic pseudogap and superconducting states. 

Although for small variations of thermodynamic variables the compressibility of the electronic system is often a dull thermodynamic quantity, close to phase transitions the compressibility can vary strongly. Correspondingly, the capacitance of an electrode built from such a material will reflect these variations in the electronic state --- specifically the density dependence of the electronic energy.
It is pointed out that the compressibility is essentially proportional to the density of states 
of the interacting electronic system. 
Consequently, the capacitance depends on spectral weight transfer and formation of coherence peaks in the momentum integrated spectral function. Both of these phenomena are typical for strongly correlated electron systems and can,
to a certain extent, be controlled by temperature and magnetic field. It is therefore expected that they can produce capacitive effects for electrodes. This prediction has to be investigated further. 

Whereas one does not expect the compressibility to vary appreciably close to the Mott transition in the single band 
Hubbard model, for the multi band case the situation is markedly different: a large interorbital charge transfer with
negative compressibility is feasible in a multiband model with at least one band being close to a Mott transition, as has been 
suggested by Liebsch for the insulator-metal transition in Sr-doped LaTiO$_3$.~\cite{Liebsch08} In such multi-band (bulk) systems, the different orbitals act like the electrodes of a capacitor with negative compressibility at one ``electrode'' and zero distance between the ``electrodes''.
Similar to these multiband correlated materials, one may expect for mixed or intermediate valence systems
a stronger dependence of the compressibilty on control parameters such as doping and temperature.

A suppression of the density of states at the Fermi surface is known to occur in disordered correlated electronic systems~\cite{Altshuler79}. Al'tshuler and Aronov identified a square root suppression of the density of states at the Fermi level with decreasing temperature as a consequence of the interference between inelastic and multiple, elastic scattering processes. Therefore we expect a corresponding decrease of the capacitance in case the surface of the electrodes consists of such a material.

\section{Capacitors with Multiple Stable or Metastable States}

\begin{figure}[b]
\centering
\includegraphics[width=1.0\columnwidth]{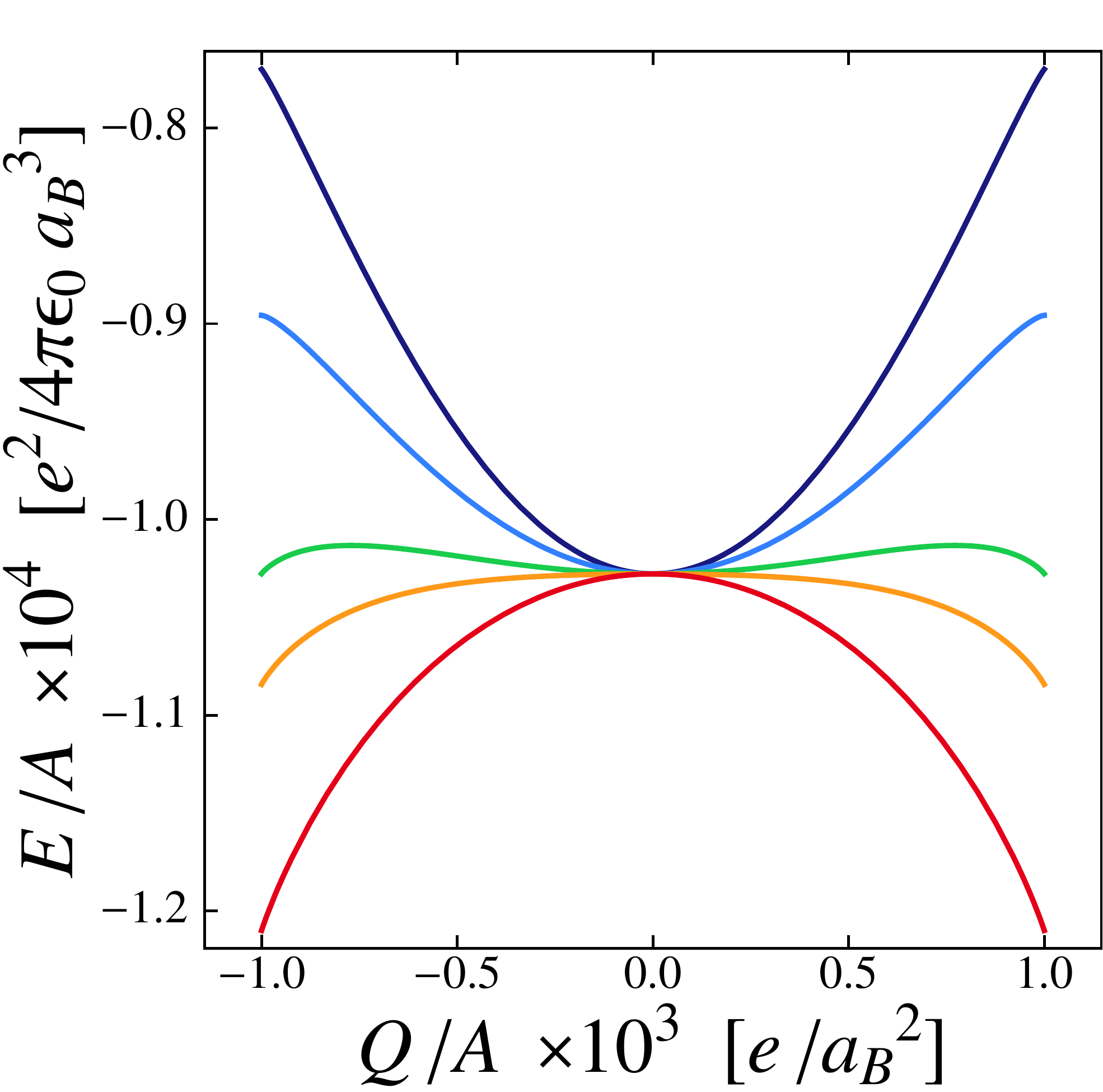}
\caption
{Energy per area of a two-dimensional, parallel-plate capacitor plotted as a function of charge per area for several spacings of the plates. The energy has been calculated for the model capacitor described in Sec.~\ref{sec:2D}. The electronic systems of the electrodes form dilute and homogenous 2D electron gases. The charge carrier density (carriers per area) is 
$10^{-3} /a_B^2 = 3.6  \times 10^{13} {\rm cm}^{-2}$, the effective mass of the carriers is $m^\star/m = 1$, 
and the effective dielectric constant in the electrodes is $\epsilon_{{\rm eff}}=1 $ for all lines. 
The effective distance between the plates $d^\star/a_B = d/(\epsilon_r\, a_B)$ is 10.0, 8.0, 6.0, 5.0, and 3.0 from top to bottom.
}
\label{Energy}
\end{figure}

\begin{figure}[t]
\centering
\includegraphics[width=1.0\columnwidth]{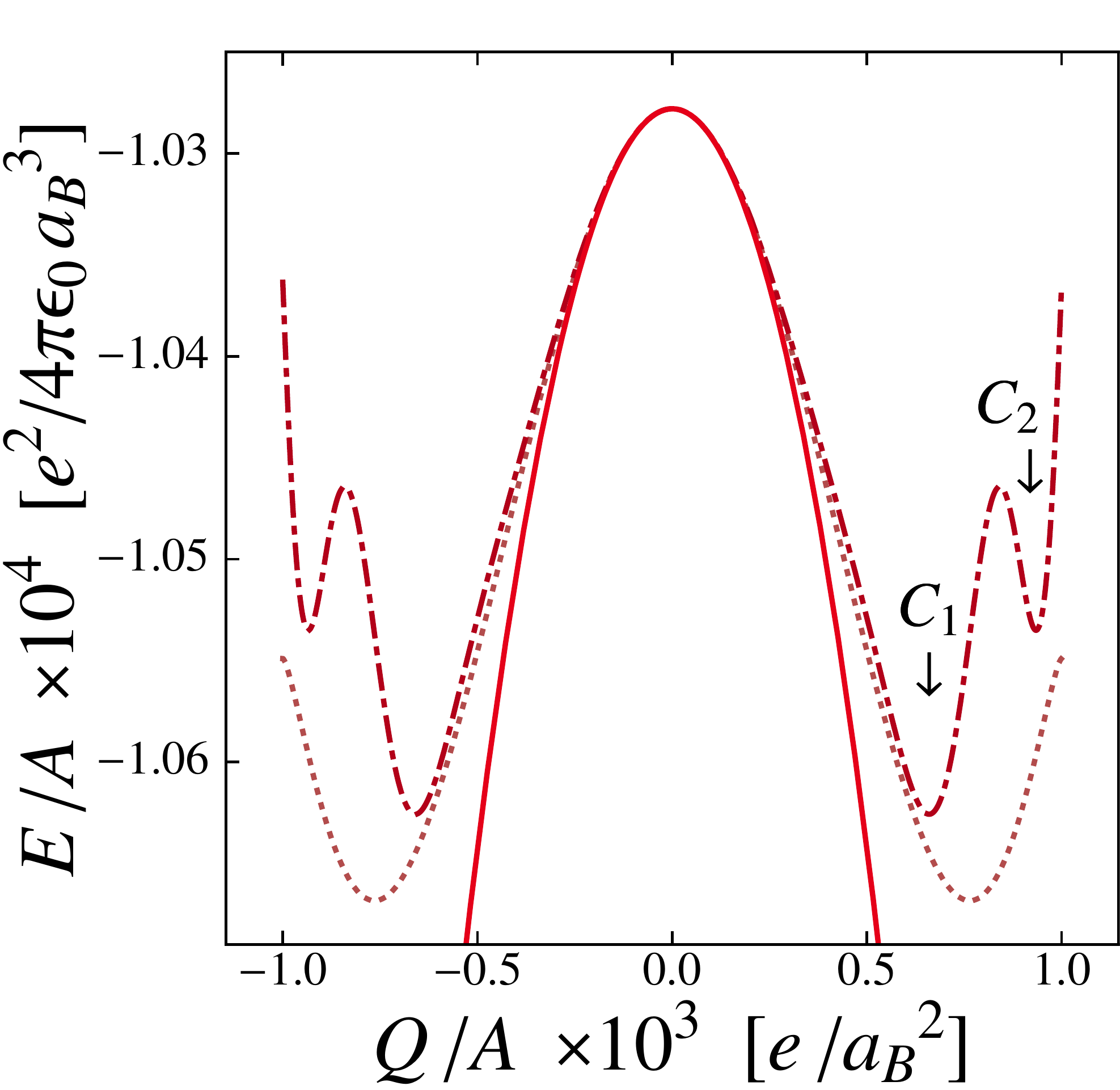}
\caption
{Sketch, illustrating several possible scenarios for the charge dependence of the energy 
of the two-dimensional, parallel-plate capacitor (compare Fig.~\ref{Energy}).
The continuous red line corresponds to the red line of Fig.~\ref{Energy} with $d^\star/a_B = 3.0$. The dotted and dash-dotted lines are tentative extrapolations for high charge densities of the order of the carrier density
$10^{-3} /a_B^2 = 3.6  \times 10^{13} {\rm cm}^{-2}$. The latter lines have not been calculated, but rather present a sketch. The capacitor with four minima has different capacities in the metastable and stable states, as reflected by the different curvatures of the characteristics in the minima. Here, $C_1$ and $C_2$ refer to the inverse of the curvature at the minima which are marked by the vertical arrows. The capacitance of such a capacitor can therefore be switched by current pulses.
}
\label{EnergyScenos}
\end{figure}

The energy functional of the capacitor is given by  Eq.~(\ref{energy}). 
The energy $E$ as a function of 
charge per area, $Q/A$, is identified from the dependence of $E$ on the charge carrier density. 
For the model capacitor (Sec.~\ref{sec:2D}), for example, the energy per area is plotted as a function of 
charge density in Fig.~\ref{Energy}.
The curvature at $Q/A=0$
represents the inverse differential capacitance $C^{-1}(Q=0)=dV/dQ$ which was 
calculated in Sec.~\ref{sec:2D}. 

Due to the negative capacitances of this model system, all electrons will accumulate on one of 
the two equivalent parallel plates, leave holes on the other plate, and thereby charge the capacitor~\cite{comment-SILC} 
(see the lower two curves in Fig.~\ref{Energy}). 
The capacitor is characterized by multiple stable states, and may be switched between these 
by small bias pulses. For
a capacitor with $d^\star/a_B = d/(\epsilon_r\, a_B)=5.0$ (the second curve from the buttom, in orange) and
an area of $A=100\, {\rm nm}^2$, the energy difference to the zero charge state is approximately 0.5~eV,
and the states will be stable aganst thermal activation. In this model, also capacitors with three stable or 
metastable states can be realized (the third curve from the buttom, in green), 
however with an energy difference of only about  50~meV.

The model is not suited to precisely describe the multiple-state properties of real capacitors, because the
nonlinear response of an inhomogeneous electron system will typically stabilize a charged state at charge 
densities that are smaller than provided by the model.
This scenario is illustrated in Fig.~\ref{EnergyScenos} (dotted curve). 
Phase transitions in electrodes with strongly correlated, inhomogeneous electron systems might even 
produce additional minima in the charge dependence of the energy functional 
(see Fig.~\ref{EnergyScenos}, dash-dotted curve).
The investigation of such multistable charge states in capacitors that only use dielectric and not ferroelectric dielectrics is considered relevant for electronic devices such as memory devices or Qbits.

\begin{figure}[b]
\centering
\includegraphics[width=0.95\columnwidth]{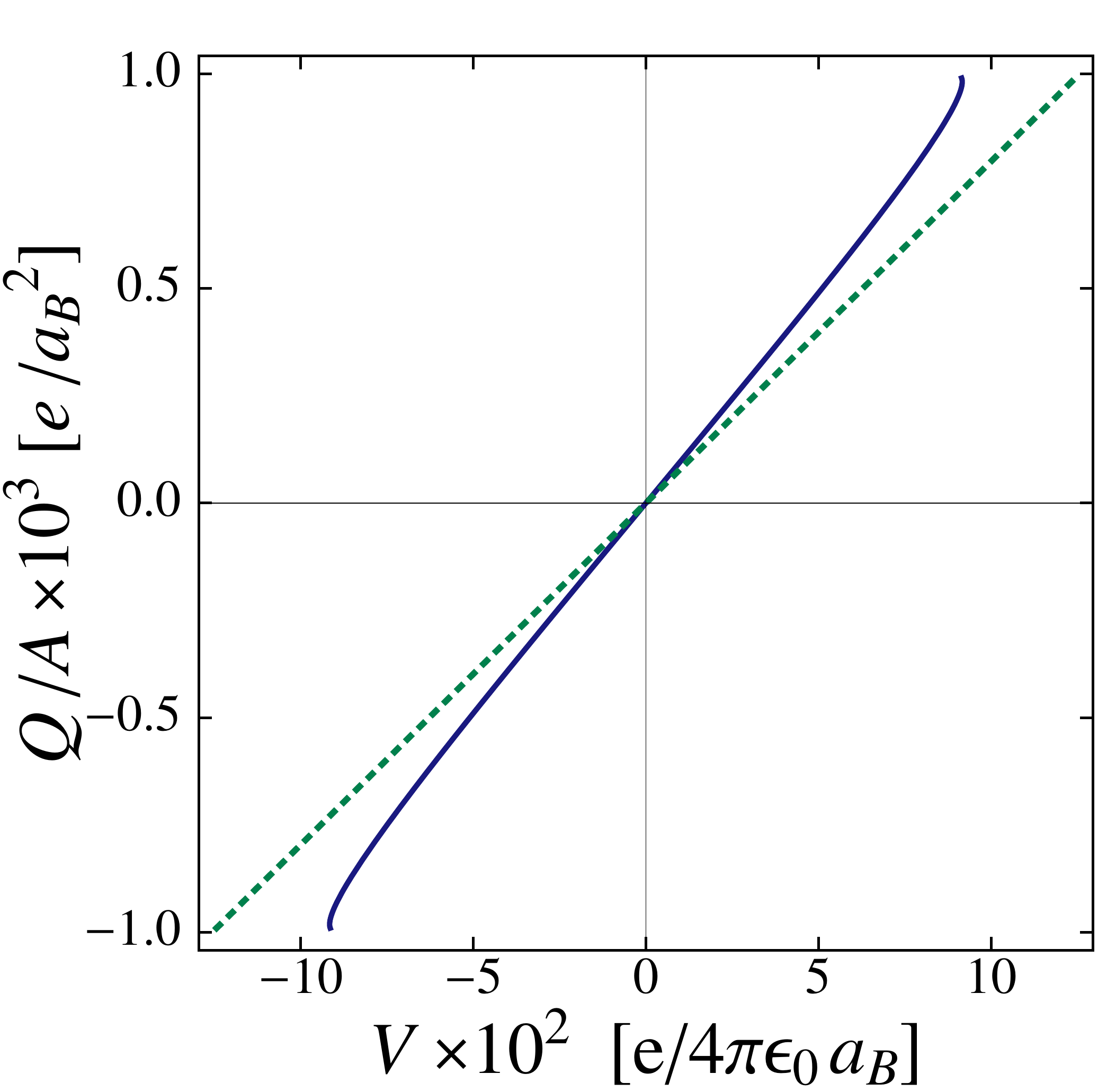}
\caption
{
Charge per area of a two-dimensional, parallel-plate capacitor plotted as a function of voltage between the plates. The charge per area $Q(V)$  has been calculated for the model capacitor described in Sec.~\ref{sec:2D}. The electronic systems of the electrodes form dilute and homogenous 2D electron gases. The charge carrier density (carriers per area) is 
$10^{-3} /a_B^2 = 3.6  \times 10^{13} {\rm cm}^{-2}$, the effective mass of the carriers is $m^\star/m = 1$, 
and the effective dielectric constant in the electrodes is $\epsilon_{{\rm eff}}=2.5 $. 
The effective distance between the plates $d^\star/a_B = d/(\epsilon_r\, a_B)$ is 10.0, 
where $\epsilon_r$ is the dielectric constant of the dielectric.
The dotted line shows the $Q(V)$-characteristic of the conventional, classical capacitor with $Q=C_{\rm geom} V $.
}
\label{QV}
\end{figure}

Capacitors with negative differential capacitance generate in the charged state electric fields in the dielectric and an electrical voltage between their electrodes. However, the charged state is typically in thermal equilibrium so that
the electro-chemical potential vanishes and no charge flows through the leads by which the two electrodes are connected. 
It is noted that while such capacitors are charged when being in one of the equilibrium states, these states are obviously stable states. The capacitors do therefore not, as all other capacitors do, loose their charge with time by discharging.
Treated as components in electronic circuits, capacitors with negative capacitances, such as the one for which the $E(Q)$ characteristic is sketched by the continuous line of Fig.~\ref{EnergyScenos}, will be characterized by the impedance $Z = - i/( \omega C)$  if operated close to the energy maximum. 
Thus, the frequency dependence of their impedance is the one of a standard capacitor. Because $C$ is negative, however, the induced phase shift has the opposite sign of the phase shift of standard capacitors, so that the phase shift corresponds to the phase shift of a standard inductance. In analogy, we suggest the possible realization of conductors with negative inductances, the phase shift of which corresponds to the phase shift of standard capacitances.

It is also amusing to consider the threshold behavior at a possible electrical breakdown of the dielectric.
When approaching breakdown, localized charge carriers in the dielectric start to get released and move to
the electrode of opposite charge. This charge redistribution shifts the electro-chemical 
potential so that an equal number of charge carriers flows through the leads. 
The original charge state is hereby stabilized and the release of
charges does therefore not seem to change the electronic state in the electrodes. 
At the same time, the change of the charge distribution in the dielectric reduces  the release of further carriers in the dielectric. The Coulomb-field of the altered charge distribution, however,  affects the energy of the electron systems of the plates and thereby indirectly alters the charge distribution on the electrodes.
Should at breakdown a conducting channel be induced in the dielectric, the same argument applies, if the channel is considered as an additional lead that connects the electrodes.
The electronic state of the electrodes is not 
dependent on this channel and the electrodes will stay charged.

\section{Voltage Dependence of the Capacitance}  

\begin{figure}[b]
\centering
\includegraphics[width=1.00\columnwidth]{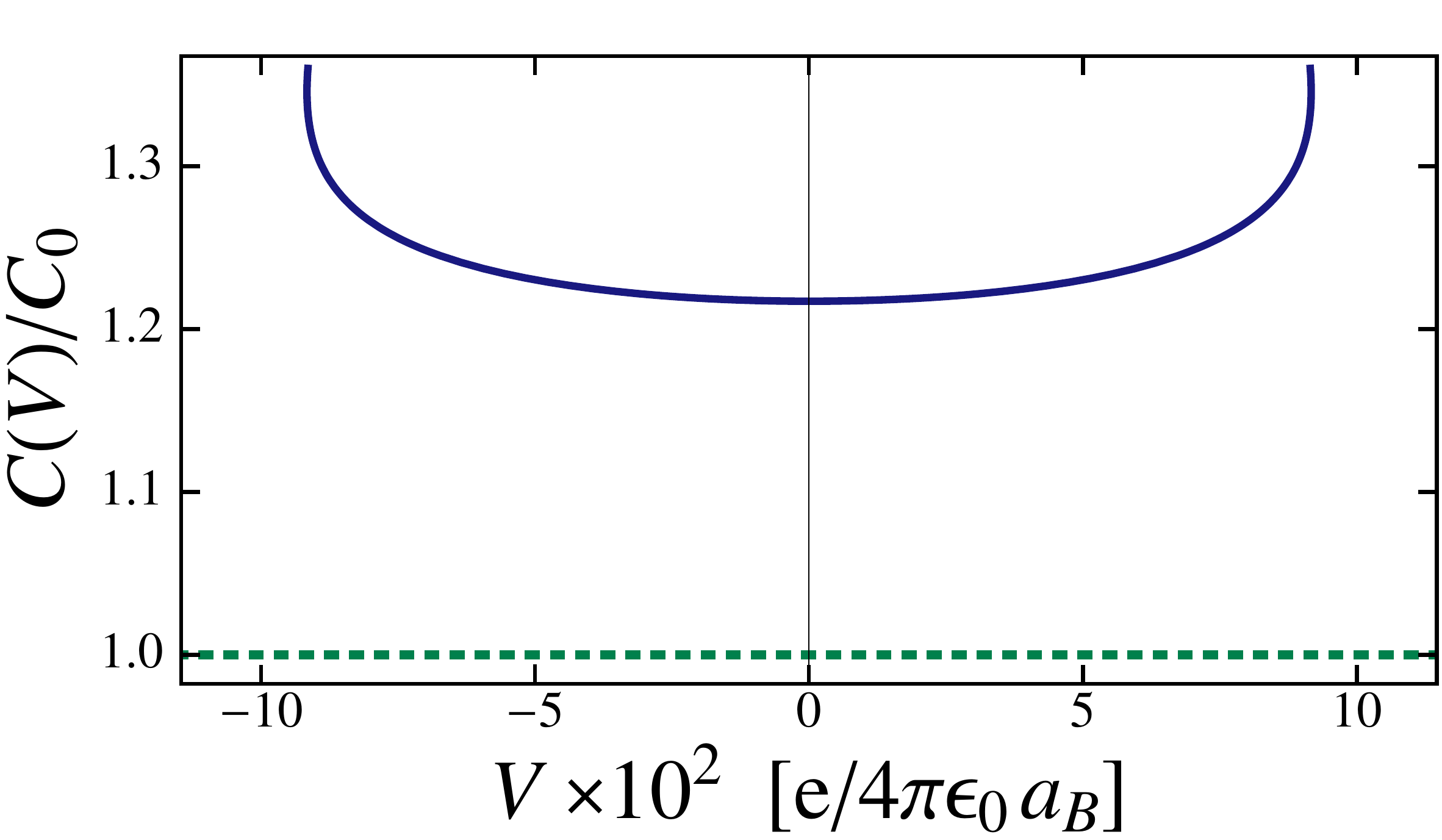}
\caption
{
Capacitance $C(V)=Q(V)/V$ of a two-dimensional, parallel-plate capacitor plotted as a function of voltage. 
The capacitance $C(V)$  has been calculated for the model capacitor described in Sec.~\ref{sec:2D}. 
The parameters are the same as in Fig.~\ref{QV}.
The capacitance is normalized to $C_0$, the capacitance of a classical capacitor of the same geometry.
}
\label{CV}
\end{figure}

\begin{figure}[t]
\centering
\hbox{\hskip4.5pt\includegraphics[width=1.00\columnwidth]{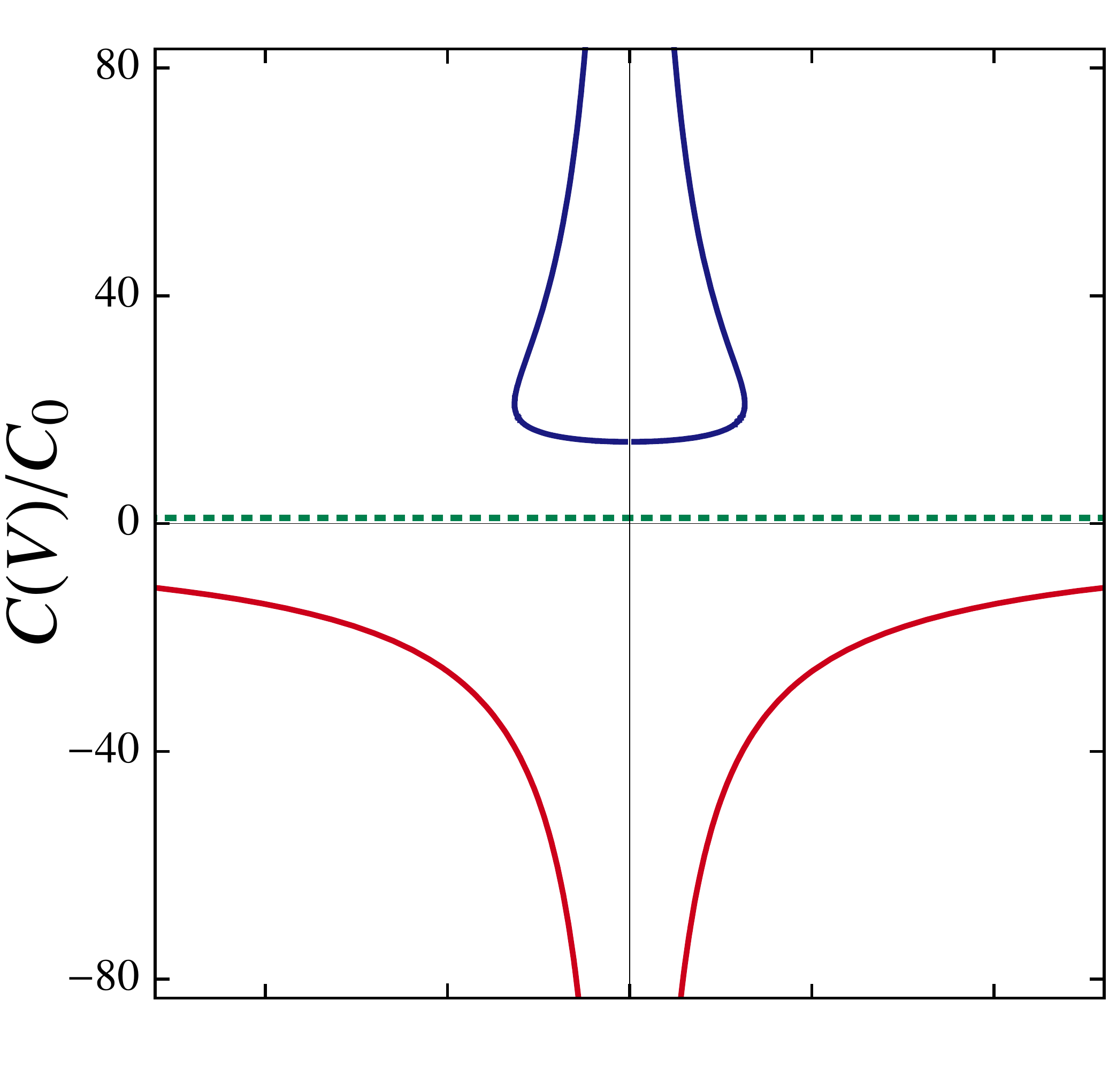}}
\vskip-1.0cm\hbox{\includegraphics[width=1.024\columnwidth]{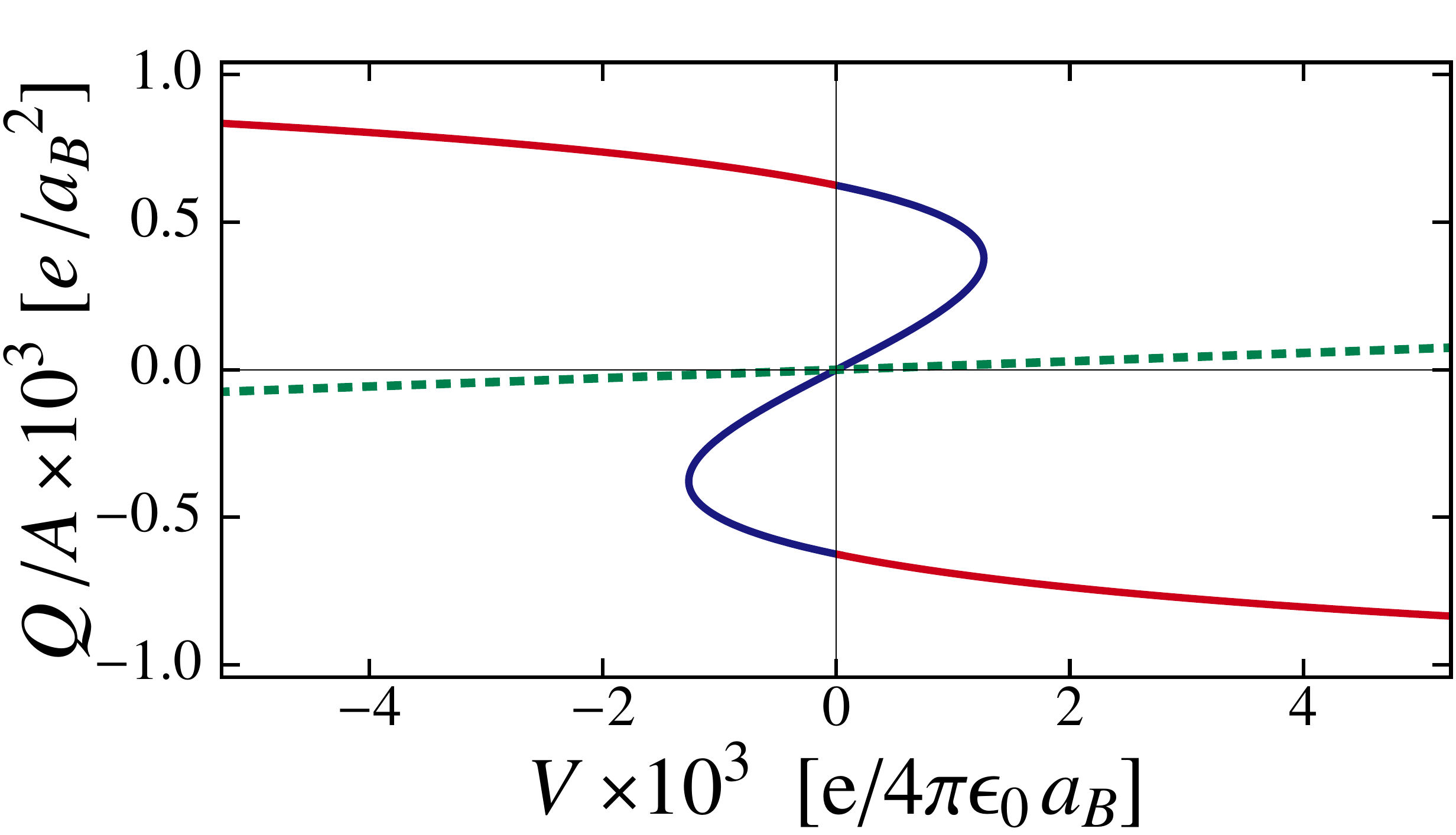}}
\caption
{
Capacitance and charge of the two-dimensional, parallel-plate capacitor of the model capacitor described in Sec.~\ref{sec:2D} plotted as a function of the voltage between the capacitor plates. 
The charge carrier density (carriers per area) is 
$10^{-3} /a_B^2 = 3.6  \times 10^{13} {\rm cm}^{-2}$, the effective mass of the carriers is $m^\star/m = 1$, 
and the effective dielectric constant in the electrodes is $\epsilon_{{\rm eff}}=1 $. 
The effective distance between the plates $d^\star/a_B = d/(\epsilon_r\, a_B)$ is 5.6.
Top panel: voltage dependence of the capacitance $C(V)=Q(V)/V$. Lower panel: voltage dependence of the charge per area $Q(V)$. In both panels, blue and red lines refer to the ranges of positive and negative capacitances, respectively. The dotted line shows the characteristics of the conventional, classical capacitor with $Q=C_{\rm geom}V $.
}
\label{QCV}
\end{figure}

In many applications it is the voltage and not the charge, with which the state of a capacitor is controlled. To calculate the $C(V)$ characteristic we consider the Legendre transform  ${\cal E}$ of the energy $E(Q)$ with respect to $Q$, and identify the total capacitance as 
\begin{equation}
\label{C(V)}
C(V)= \frac{Q(V)}{V}
\end{equation}
As example we take again the model system of  Sec.~\ref{sec:2D} with parallel, two-dimensional plates comprising the capacitor. 

The calculations reveal that already in the regime of moderate effective distance of the capacitor plates, deviations from the conventional characteristics of a textbook capacitor are manifest:
in Fig.~\ref{QV} we display $Q(V)$ which is expected to be linear with $Q=C_{\rm geom} V $ in the textbook case or for sufficiently large $d^\star =d/\epsilon_r$ (green dotted line). For $d^\star=10 a_B$ and $\epsilon_{{\rm eff}}=2.5 $ the slope at zero voltage is steeper and $Q(V)$ deviates from a linear behavior, most prominently for the strongly charged capacitor. The corresponding voltage dependent  capacitance $C(V)$ is shown in Fig.~\ref{CV}. With increasing voltage the capacitance increases because the capacitance is a function of the charge in the plates. The higher the charge in the plates the higher the capacitance which results in a noticeable upturn of $C(V)$.

This feedback of increasing capacitance with increasing charge may be so strong for nanoscopic distances that the conventional $Q(V)$ and $C(V)$ characteristic is completely modified (see Fig.~\ref{QCV}). For example, for $d^\star=5.6 a_B$ (and $\epsilon_{{\rm eff}}=1$) we operate in a regime where the capacitor self-charges once a sufficiently high charge has been accumulated. In that case charge is transfered at vanishing energy expense (cf.~Fig.~\ref{Energy}, where $V(Q)=dE/dQ=0$ for a distinct, finite $Q$ in the green, middle curve). The voltage is not zero in the fully charged state (Fig.~\ref{QCV}) because this is not a thermodynamic state with a well-defined minimum in the energy $E(Q)$ --- as compared to the tentative capacitors of Fig.~\ref{EnergyScenos} which display minima in $E(Q)$.

The capacitance $C(V)$ of the model capacitor (top panel of Fig.~\ref{QCV}) may have multiple values, depending on the charge state. It is conceivable to switch between a positive value of $C(V)$ (blue curve in Fig.~\ref{QCV}) and a negative value (red curve in Fig.~\ref{QCV}), using charge pulses to switch the capacitor in and out of the ``self-charged'' state.

\section{Specific Examples}   

Magnetocapacitive effects have been investigated for several decades. A capacitive measurement, in which Landau levels in a 2D electron gas of a MOSFET were observed, was reported in 
Refs.~\onlinecite{Kaplit68,Zemel74,Smith85,Mosser86,Ashoori92}. More recently,  the density of states in the fractional quantum Hall regime has been investigated by magnetocapacitive 
measurements.~\cite{Smith86,Eisenstein92} 
A charge controlled transition of an electrode into a magnetic state would also yield large magnetocapacitive effects.
Such a transition is conceivable in correlated electronic systems, for example, in the manganites. A positive magnetocapacitance has indeed been observed in La$_{0.7}$Sr$_{0.3}$MnO$_{3-\delta}$-titanate junctions~\cite{Hwang05} which contrasts to the negative magnetocapacitance of Pd-AlO$_x$-Al thin-film structures.~\cite{Hebard03}
Whereas the latter is the predicted behavior for a paramagnetic metal in which the Zeeman-split narrow $d$-bands cause the magnetocapacitance, the positive magnetoresistance has been related to strong electronic correlations. 

Artificial materials that, for example, consist of layered structures such as superlattices, of which the capacitance  is influenced by non-geometrical capacitances are predicted to be characterized by unusual properties. Of particular relevance are superlattices with a  geometry chosen that the capacitances between the layers are negative (see Sec.~\ref{sec:e-gas}) so that the layers self-charge,  and in which a part of the layers consist of materials with magnetic phase transitions. Applied electric fields, on the one hand, change the carrier density of the magnetic layers and therefore their magnetic properties. Application of magnetic fields, on the other hand, alters the correlation energies and therefore the capacitance and electric fields of the system.  Thus, such artificial materials are candidates for strongly coupled, room temperature multiferroics.

\subsection{Electrodes with 2D Electron Systems in Transition Metal Oxide Heterostructures}

The possibility to control the capacitance of capacitors by adding a thin, conducting sheet to an electrode opens the question how large a capacitance of a capacitor is expected that uses  as an electrode a two-dimensional electron gas generated at an oxide interface~\cite{Ohtomo04,Mannhart08}. Indeed, capacitors based on two-dimensional electron gases in oxide heterostructures are expected to have unique and possibly technically useful properties.  

The current model systems for such interfaces are systems like the $n$-doped LAO/STO or LVO/STO heterostructures. The following discussion will refer to such interfaces, although future multilayers that do not imply sophisticated epitaxial growth may be preferable for applications. For simplicity, we neglect strong correlation effects possibly present at such interfaces. 

As shown by Eqs.~(\ref{cap-geom},\ref{cap-2D-kin},\ref{cap-2D-xc},\ref{cap-2D-corr}), the capacitance 
(Eq.~(\ref{capacitance})) of a capacitor that uses such an electron gas is a function of $r_s$, of the effective mass of the interface charge carriers $m^\star$, and of the effective dielectric constant 
$\epsilon_{\rm eff}$ in the interface sheet. 

Up to now, such oxide interfaces have been prepared with sheet carrier densities of the order of $n_i \sim 5\times10^{12}-5\times10^{14}/{\rm cm}^2$, corresponding to bare $r_s$-values ranging from 50 to 5.  As Fig.~\ref{2Dcontour} shows for the example of a parallel-plate capacitor, that comprises a 1.5~nm thick HfO$_2$ layer as dielectric (with $\epsilon_r= 23$) and an electron gas with $r_s=7.0$ in each of the electrodes, the capacitances can be negative for realistic sets of $m^*$ and $\epsilon_{\rm eff}$ values. Moreover, the capacitance  is enhanced for a broad range of $\epsilon_{\rm eff}$ and $m^\star$.

\subsection{Other Systems}

$1/C_{{\rm kin}}$ may be strongly reduced at van Hove singularities. Conversely, the capacitance of condensators built with electrode materials such as graphene or other materials with a small density of states, will be reduced because then $1/C_{\rm kin}$ dominates $1/C_{\rm geom}$. In that case, the small density of states determines the charging of the capacitance, and the charge is smaller than expected from $C_{\rm geom}$.

Recently, the quantum capacitance of coaxially gated carbon nanotubes~\cite{Latessa05} has been calculated using a Green's function density functional tight-binding approach which is based on a local density approximation (LDA) of the exchange correlation functional. This quantum capacitance includes the terms $C_{\rm kin}$ and $C_{\rm x}$ which are evaluated for the quasi one-dimensional nanotube.
The behavior of this capacitance (see Fig.~3 in Ref.~\onlinecite{Latessa05}) is consistent with our results for the capacitance of 2D or 3D capacitors. Specifically, in the regime of small carrier densities overscreening of the gate field was observed which corresponds to a negative capacitance of the carbon nanotube. The capacitance of carbon nanotube has been measured by a group from Cornell University~\cite{Ilani06} and, in fact, the measurements suggest the existence of a negative contribution to the total capacitance.

A negative electronic compressibility is also expected close to structural transitions. Commonly, this signifies that the electronic system is in a phase separated state or undergoes charge disproportionation. In many cases, these situations are associated with a metal-insulator transition. If structural transitions can achieve a functional quality for a controlled electronic transition in the electrodes of a capacitor has to be investigated.

\section{Summary, Conclusions and Perspectives}

Considering the complete energy of capacitors, which includes the Coulomb energy and the energy of the electron systems of the capacitor plates, we have evaluated the capacitance of capacitors. We find that in general the capacitance is given by Eq.~(\ref{capacitance}) and not by Eq.~(\ref{capacity}). 

As shown, the capacitance of a capacitor is described by the series connection of the Coulomb capacitance, the kinetic capacitance of the electrons, the exchange capacitance, the correlation capacitance, and the capacitance which results from the interaction of electrons with non charge-carrier degrees of freedom. Because for a free electron gas the exchange capacitance and the correlation capacitance are negative and the electron-phonon interaction may also result in a negative value, the total capacitance may be smaller or larger than the Coulomb capacitance, and may even reach negative values.

Our analysis shows that the capacitance is dominated by the field capacitance for the case that the field energy exceeds the energies of the electron system, but that the electron energies are significant if the thickness of the dielectric layer divided by its dielectric constant is comparable to the Bohr radius. This applies independent of the lateral size of the capacitor. The gate capacitances of modern MOSFETs are reaching this regime.

Eq.~(\ref{capacitance}) is of general importance to calculate the capacitances of capacitors. Due to their broad applicability, we have specifically provided the detailed calculations of the capacitance of parallel-plate capacitors with electrodes that contain two-dimensional and three-dimensional, homogenous electron systems.

\begin{figure}[b]
\centering
\hskip -1.8cm\includegraphics[width=0.75\columnwidth]{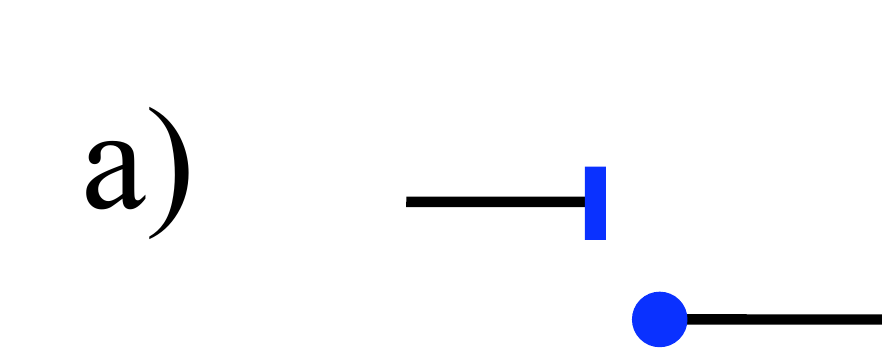}
\vskip 1cm
\includegraphics[width=0.95\columnwidth]{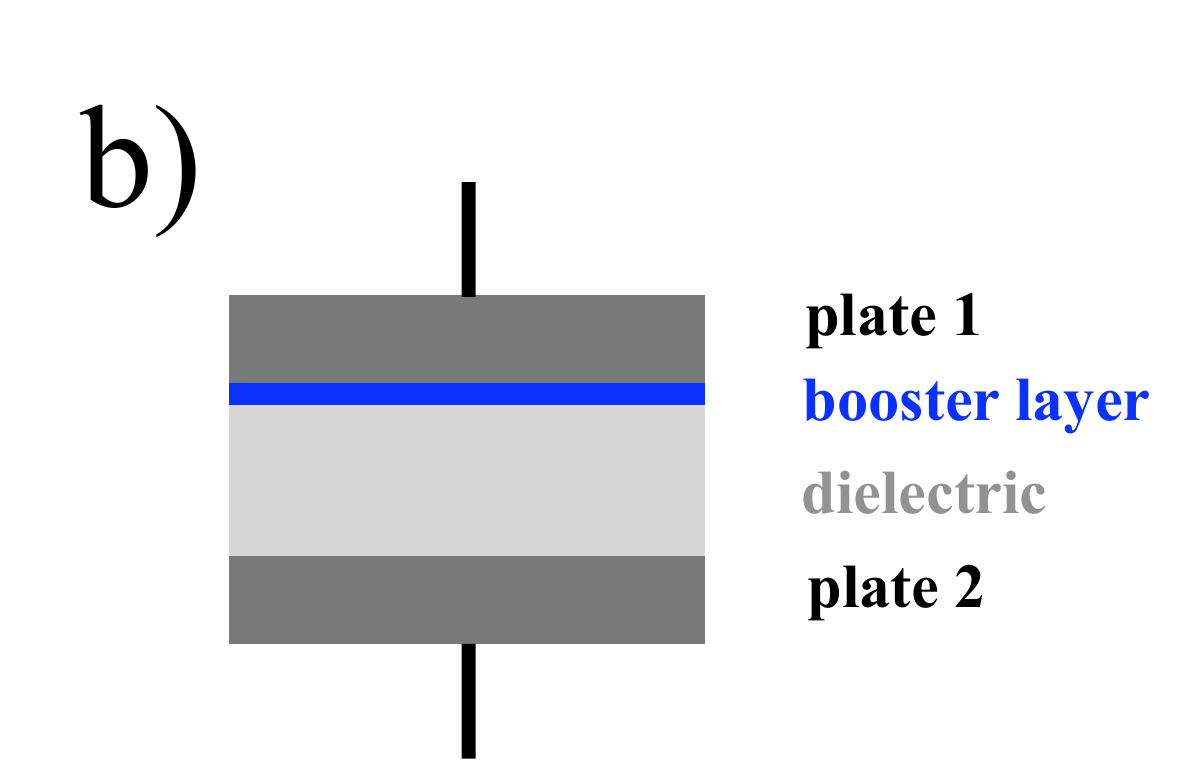}
\caption
{Sketch, illustrating options to integrate materials into capacitors to achieve capacitances that deviate strongly from the geometric capacitance. Panel (a) sketches 
a situation where electrodes may cause a strong variaton of the capacitance due
to their contributions in $C^{-1}_{\rm kin}$, $C^{-1}_{\rm x}$, $C^{-1}_{\rm c}$ or $C^{-1}_{\rm el-ncc}$ so 
that a large total capacitance is obtained without using a parallel-plate configuration.. Panel (b)
sketches a thin, conducting sheet (booster layer) of a compound with appropriate electron energies integrated into a standard capacitor. This sheet can considerably enhance or reduce the capacitance. The tickness of the sheet may be as small as the electric screening length in the compound.
}
\label{Configs}
\end{figure}

It is concluded that the capacitance of capacitors can be significantly enhanced or decreased by altering the material of the electrodes. It suffices to introduce thin, conducting sheets of a compound (see Fig.~\ref{Configs}b) with the appropriate electron energies to drastically alter the capacitance. 

Therefore, besides optimizing the thickness and the dielectric properties of the dielectric layer, choosing the appropriate materials of the electrode surfaces provides an independent access to optimizing the capacitance of capacitors. Materials of interest include materials with small electron densities, such as interfaces in oxide heterostructures, low-electron density metals, materials with strong electronic correlations such as transition metal oxides, or graphene-based systems.

Because the capacitance of such capacitors is controlled by the energy of the electron system of the plates rather than by the field energy, the standard parallel-plate configuration is not needed to obtain large capacitances or field strengths (for example, in Fig.~\ref{Configs}a all blue surfaces contribute to the capacitance $1/C -1/C_{\rm geom}$ in Eq.~(\ref{capacitance})). Rather, the surface areas of the electrodes are important.  We note that the desired modification of the capacitance does not require that the electrodes consist entirely of  strongly correlated compounds. It is sufficient to coat the active surface of an electrode with a thin layer of a correlated material, because the net charges are repelled from the interiors of the electrodes.

We emphasize that ${\epsilon_{{\rm eff}}}$ is a property of the electrode materials and, specifically, it relates to the properties of the materials at their surfaces. $ {\epsilon_{{\rm eff}}}$ therefore can be tuned by coating the electrodes with thin layers of desired materials and then is given by the dielectric constant of the coating material. It accounts for the polarizability of the ionic cores and also includes the short-range screening processes from excitations in the adjacent material layers, but not the contributions from the 
electron gas itself.  The effective dielectric constant ${\epsilon_{{\rm eff}}}$ is, beyond linear response, also dependent on the charge density. Such a dependence induces feedback effects which alter the capacitance.
While one may consider to estimate ${\epsilon_{{\rm eff}}}$ from an LDA evaluation for the specific heterostructure, 
there is no simple scheme to calculate ${\epsilon_{{\rm eff}}}$. Lacking such a simple scheme,  
we calculated the capacitance of a simple ``model capacitor'' for several values of ${\epsilon_{{\rm eff}}}$.

\begin{figure}[t]
\centering
\includegraphics[width=0.6\columnwidth]{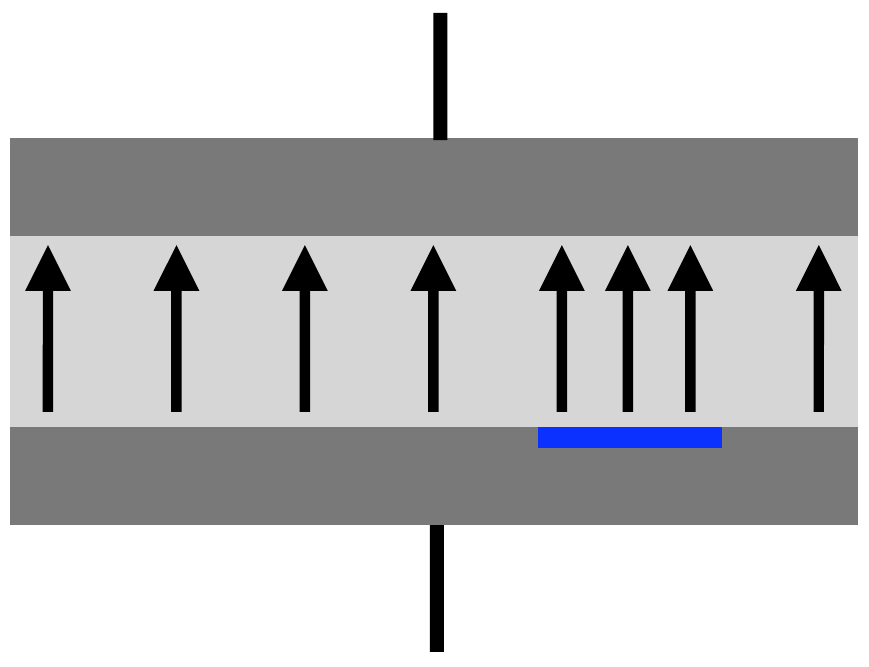}
\caption
{Sketch to illustrate the possibility of electric field enhancement due to the integration of a material with high capacitance into a plate of a capacitor. The material is introduced in a partial coating (blue stripes) of the capacitor plates (dark gray rectangles).
}
\label{FieldFocus}
\end{figure}

Using optimized capacitor configurations, it is in principle possible to build capacitors that are characterized by a negative capacitance for zero applied voltage. Such capacitors charge themselves to generate finite electric fields and, although they do not contain any ferroelectric component, show parallels to ferroelectric devices. It is obvious, that for device applications novel possibilities are opened by adding truly ferroelectric layers to structures with unconventional capacities. Intriguing characteristics arise if the capacitances are negative.  

In the case of negative compressibility, adding or removing an electron from an electrode
involves less energy than the standard work function or electron affinity. It is intriguing to regard the behavior of the electrons in a single electrode or on the two electrodes of a capacitor, in case the compressibility of at least 
one electron system is negative. In a capacitor 
that self charges, for example, into a positively and negatively charged plate, the electrons and the holes in the two
plates interact across the dielectric by their Coulomb field. It seems possible that electron-hole pairs result that 
form a coherent state, and, for example modify pairing in a superconducting electrode. Such effects may be 
particularly useful in case the plate capacitor is designed as a transmission line. We propose that such a superconducting
state may even exist if none of the electrodes is superconducting by itself. Even for electron systems with positive 
compressibilities, the Coulomb energies caused by the presence of a second plate will 
lower or enhance the $T_c$ 
of a superconducting electrode. Therefore, modifying Coulomb energies by using capacitor-type configurations, the 
electrodes of which may even consist of the same materials, offers new possibilities to realize superconductors. 

Also, because the capacitance can be controlled by the electronic system of the plates, which may be strongly correlated electron systems that are sensitive to magnetic fields, materials may be designed from capacitor-derived multilayer configurations that yield large multiferroic coupling at room temperature. 

Capacitors for which the non-geometrical capacitances are important, and in which one plate 
has a non-homogeneous composition, are characterized by inhomogeneous and unusual 
electric field distributions.
The charge distribution of such an electrode is modified and differs from the charge distribution generated
in the corresponding standard configuration. Consequently, in the
presence of an electric field, the distribution of the field lines is accordingly
altered. As illustrated in Fig.~\ref{FieldFocus}, electric field lines can, for example, be
focused onto desired areas to extract charges. Or, where
desirable, areas can be shielded from electric fields, which may be important to
prevent electric breakthrough phenomena. Thus, inhomogeneous electrodes 
allow to generate desirable distributions of charge
densities and electric fields without altering the overall geometrical configuration
of a device.


\begin{acknowledgments}
We would like to thank Yu.~S.~Barash for his careful reading of the manuscript and for bringing several important references to our attention. It has been very helpful that S.~Graser kindly double-checked relations and plots.
Illuminating discussions with B.~Batlogg, M.~Breit\-schaft, P.~Chaudhari, V.~Eyert, P.~J.~Hirschfeld, R.~Jany, A.~Kampf, C.~Richter, D.~Scalapino, D.~G.~Schlom,  J.~Schubert, S.~Thiel, and P.~W{\"o}lfle are gratefully acknowledged.
\end{acknowledgments}


\begin{thebibliography}{99}
\bibitem{commentT} We have fabricated tunnel junction cosisting of SrTiO$_3$-LaAlO$_3$-Au trilayers and measured the tunnel currents. To obtain significant tunnel currents, the LaAlO$_3$ layer has to be at most 4 unit cells thick, i.e., 1.6~nm. The changes of the capacitance described in the paper are expected to occur also at larger thicknesses. 
Even for a 1.6~nm thick barrier, the effects of the tunnel currents on the energy of the capacitances is small.
\bibitem{commentV} For clarity, we point out that the expression ``external potential'' used in characterizing $E_{{\rm ext},i}$ does not refer to the electrochemical voltage applied to the capacitor plates by an external voltage source.  The electrochemical potential is the thermodynamic conjugate variable to the total charge $Q$ and is derived from the total energy through $V(Q) = dE(Q)/dQ$,  i.e., from the sum of all energy terms in Eq.~(\ref{energy}).
\bibitem{comment0} We use the term ``homogeneous electron system'' in the following, standard sense: a
homogeneous electron system is an electron
system with a positive background charge-density $n=N/{\cal V}$ 
that is uniformly spread through the volume ${\cal V}$, where
$N$ is the number of electrons in the neutral system. The positive background is supposed to be inelastic. 
The charged capacitor carries the additional electronic charge $\pm Q$ on each of the plates.
\bibitem{commentcap}  This relation is obtained from the expansion of $E'(Q)=E(Q)-VQ$ up to quadratic order in $Q$, i.e., from $E'(Q)=E_0 + Q \Delta\Phi/e + Q^2/(2C) - VQ$ and the thermodynamic relation $dE(Q)/dQ =V$. 
The energy function
 is equivalently written in the form $E'(Q)=E_0 - \frac{1}{2}C (\Delta\Phi/e)^2 + (Q-Q_{\rm s})^2/(2C) - VQ$ with 
 $Q_{\rm s}=-C\Delta\Phi/e$, which directly identifies $C$ (Eq.~(\ref{capacitance})) as the capacitance.
 Alternatively, one may define $C'=Q(V)/V$ and derive the relations
 $C'^{-1}=C^{-1}+(\Delta\Phi/e)/Q$ and $C'=C(1-(\Delta\Phi/e)/V)$ with $C^{-1}$ from Eq.~(\ref{capacitance}).
\bibitem{comment2Dphases} For a homogeneous 2D electron gas, the compressibility becomes negative at $r_s\simeq 2.03$ for the paramagnetic state,~\cite{Tanatar89,Bulutay02} and at $r_s\simeq 3.07$ for the fully 
spin-polarized state.~\cite{Bulutay02}  There are indications for a ferromagnetic ground 
state beyond $r_s\simeq 30$,~\cite{Varsano01} but a transition of the ground state to a 
ferromagnetic state has not yet been established in 2D.~\cite{Tanatar89,Bulutay02} 
In any case, the energy difference is only in the milli\-Rydberg range. 
The Wigner crystallization~\cite{Wigner38} into a triangular crystal is estimated to occur at 
$r_s = 37\pm 5$.~\cite{Tanatar89} However, at $r_s\simeq 40$
all phases have approximately equal energies.~\cite{Tanatar89}
\bibitem{comment3Dphases} For a homogeneous 3D electron gas, the compressibility becomes negative at $r_s\simeq 5.4$ for the paramagnetic state.~\cite{Ceperley78}   A transition of the paramagnetic ground state to a 
ferromagnetic state has been predicted in 3D.~\cite{Bloch29,Ceperley78} The values of $r_s$, at which the transition
takes place, are still difficult to estimate. Recent evaluations suggest a continuous transition over the range $20\pm 5\leq r_s\leq 40\pm 5$ with a fully polarized state beyond $r_s\simeq 40$,~\cite{Ortiz99} or rather a second order transition to a partially polarized phase at $r_s = 50\pm 2$ with full polarization at $r_s\simeq 100$.~\cite{Zong02} 
The transition to the ferromagnetic bcc Wigner crystal is found to be first order at  $r_s =65\pm 10$,~\cite{Ortiz99} or 
at  $r_s =100\pm 20$.~\cite{Alder80,Zong02}
\bibitem{comment2} A multi-band calculation also has to include modified exchange and correlation terms, an important subtlety that we do not address in this paper.
\bibitem{comment3} We interpret the relation ``$\alpha_4\; h_0=-0.2760$'' in 
        Table~IV of Ref.~\onlinecite{Ceperley78} to be ``$\alpha_4/h_0=-0.2760$''; following the possible misprint, the correlation energy would be negative.
\bibitem{comment1} Of course, for  $\epsilon_{r}=1$ it is senseless to discuss a transition to 
negative capacitance regime as $d$ cannot be as small as a Bohr radius.
\bibitem{comment4} On first thought, one may consider to include 
also exchange and correlation processes in the screening lengths $\lambda_i(r_s)$. The static 
screening length is determined by $e^2\lambda_i^2 =\epsilon_0{\partial\mu_i}/{\partial n_i}
 =\epsilon_0{\varkappa_i^{-1}}/{n_i^2} $ which results in a vanishing $\lambda_i$ at the transition
into a negative compressibility state so that $\lambda_i$ attains a complex value  for even larger $r_s$. 
However,  $\lambda_i$ does not characterize the width of the layer of induced charge 
  at the surface of the electrodes. A more
 appropriate quantity $\lambda_s/r_s a_B$ was defined by 
 Schakel (Eq.~(20) in Ref.~\onlinecite{Schakel01}). This length coincides
 with the Thomas-Fermi screening length $\lambda_i^{{\rm (TF)}}$ for small $r_s$ and decreases
 monotonically in the entire range $0<r_s<15$ roughly with $r_s^{-1/2}$. We 
 approximate the width by $\lambda_i^{{\rm (TF)}}$ in order to discuss the evolution of the capacitance with
 increasing $r_s$ at least in a qualitative way.
\bibitem{comment-SILC} As discussed in the introductory paragraphs of Section~\ref{sec:e-gas}, the ground state for a symmetric capacitor is characterized by a spontaneous interlayer phase coherence (SILC) in the absence of self-charging. However, the coherence may be lost, if the symmetry is broken, for example, by a finite bias or asymmetric configurations. Here,
we consider the unpolarized homogeneous electron gas on a parallel plate capacitor in the absence of SILC as a ``model system''
in the expectation that it shares some general features with capacitors comprising strongly correlated electron systems in the presence of a bias.
\end{thebibliography}
\end{document}